\documentclass[10pt,journal,compsoc]{IEEEtran}
\ifCLASSOPTIONcompsoc
  \usepackage[nocompress]{cite}
\else
  \usepackage{cite}
\fi
\ifCLASSINFOpdf
\else
\fi
\usepackage{cite}
\usepackage{amsmath,amssymb,amsfonts}
\usepackage{algorithmic}
\usepackage{graphicx}
\usepackage{textcomp}
\usepackage{xcolor}
\def\BibTeX{{\rm B\kern-.05em{\sc i\kern-.025em b}\kern-.08em
    T\kern-.1667em\lower.7ex\hbox{E}\kern-.125emX}}

\usepackage{multirow}
\usepackage{subcaption}
\usepackage{algorithm}

\begin{document}
\title{Bitcoin Transaction Forecasting with Deep Network Representation Learning}

\author{Wenqi~Wei,~\IEEEmembership{Student Member,~IEEE,}
        Qi~Zhang,~\IEEEmembership{Member,~IEEE,}
        and~Ling~Liu,~\IEEEmembership{Fellow,~IEEE}% <-this % stops a space
\IEEEcompsocitemizethanks{\IEEEcompsocthanksitem Wenqi~Wei and Ling~liu are with the School
of Computer Science, Georgia Institute of Technology, Atlanta,
GA, 30332.\protect\\
Qi~Zhang is with IBM T. J. Watson Research Center, Yorktown Heights, NY, 10598. \protect\\
E-mail: wenqiwei@gatech.edu, q.zhang@ibm.com, ling.liu@cc.gatech.edu
\IEEEcompsocthanksitem The work is done during the first author's internship at IBM T. J. Watson Research Center.}% <-this % stops an unwanted space
\thanks{Manuscript received xxxx xx, xxxx; revised xxxxx xx, xxxx.}}


\markboth{IEEE Transactions on Emerging Topics in Computing,~Vol.~xx, No.~x, xxxx~20xx}%
{Shell \MakeLowercase{\textit{et al.}}: Bare Demo of IEEEtran.cls for Computer Society Journals}



\IEEEtitleabstractindextext{%
\begin{abstract}
Bitcoin and its decentralized computing paradigm for digital currency trading are one of the most disruptive technology in the 21st century. This paper presents a novel approach to developing a Bitcoin transaction forecast model, DLForecast, by leveraging deep neural networks for learning Bitcoin transaction network representations. DLForecast makes three original contributions. First, we explore three interesting properties between Bitcoin transaction accounts: topological connectivity pattern of Bitcoin accounts, transaction amount pattern, and transaction dynamics. Second, we construct a time-decaying reachability graph and a time-decaying transaction pattern graph, aiming at capturing different types of spatial-temporal Bitcoin transaction patterns. Third, we employ node embedding on both graphs and develop a Bitcoin transaction forecasting system between user accounts based on historical transactions with built-in time-decaying factor. To maintain an effective transaction forecasting performance, we leverage the multiplicative model update (MMU) ensemble to combine prediction models built on different transaction features extracted from each corresponding Bitcoin transaction graph. Evaluated on real-world Bitcoin transaction data, we show that our spatial-temporal forecasting model is efficient with fast runtime and effective with forecasting accuracy over 60\% and improves the prediction performance by 50\% when compared to forecasting model built on the static graph baseline.
\end{abstract}

}

\maketitle

\IEEEdisplaynontitleabstractindextext

% \fi
\IEEEpeerreviewmaketitle

\IEEEraisesectionheading{\section{Introduction}\label{sec:introduction}}

\IEEEPARstart{L}{aunched} in 2009, Bitcoin is the first successful decentralized cryptocurrency system with a number of unique capabilities~\cite{nakamoto2008peer}. First, it allows users to create accounts and transact with one another on the Bitcoin peer-to-peer network in a decentralized fashion. There is no central authority that oversees the cash flow within the system. Second, it uses the Blockchain technology for secure computing without centralized authority in an open networked system. A Blockchain is a distributed database, which logs an evolving list of transaction records by organizing them into a hierarchical chain of blocks. The Blockchain is created and maintained using a peer-to-peer overlay network and secured through intelligent and decentralized utilization of cryptography with crowd computing~\cite{zhang2019security}.
Third, it employs a proof-of-work consensus protocol to verify and authenticate the transactions that are carried out in the network. Bitcoin is becoming increasingly popular and is widely recognized as the first successful example of the cryptocurrency economy~\cite{barber2012bitter,eyal2018majority}.
Bitcoin transactions have made publicly available since its inception. Most existing research efforts have centered primarily on mining the statistical characteristics of the Bitcoin transactions. We argue that it is also important, though more challenging, if we can analyze the Bitcoin transactions collected to date to extract the distinctive characteristics of Bitcoin transactions and build Bitcoin transaction inference models for transaction forecasting, transaction tracking, and user identification, to name a few. One way to learn the interesting transaction patterns is to model the Bitcoin network as a big graph with accounts (or nodes) in a Bitcoin network as vertices and transactions conducted between two accounts as the edge between two Bitcoin accounts (nodes).

In this paper, we present DLForecast, a Bitcoin transaction forecasting system, by leveraging deep network representation learning. Our goal is to  predict transaction relationships among accounts on the Bitcoin network. Example usage of such a forecasting system can be transaction pattern discovery, fraud detection, account activity prediction, and so forth.
One approach to achieving our goal is to utilize a deep neural network (DNN) to learn important hidden features among transactions on the Bitcoin transaction graph, related accounts, transaction amounts, and temporal and spatial transaction properties. The development of our transaction forecasting DNN model consists of three main tasks. First, we need to  extract observable features from a Bitcoin transaction dataset. There are three main challenges for Bitcoin transaction feature extraction: (1) As of October $14^{\text{th}}$, 2019, there are more than 464,814,264 transactions on 599,446 blocks, making the Bitcoin transaction a large network to process. (2) Some of the transaction patterns in the present days are quite different from those of 5 years or 10 years ago. How to capture the up-to-date transaction patterns for accurate analysis and prediction on demand is a challenging problem. (3) Bitcoin transaction addresses (accounts) have a short life span, and those transactions happened in the past will have a very limited impact on future transactions, and such influence also decays over time. For example, a transaction happened 8 years ago often has a negligible influence on the transaction patterns today. Thus, it is also critical to ``forget'' and to ``live in the moment".
Motivated by these challenges, we extract observable features of Bitcoin transactions by exploring spatiotemporal information in the data. By statistically analyzing the address connectivity pattern and the transaction Bitcoin amount pattern, we build the time-decayed reachability graph to represent the inter-account transaction reachability, and  the time-decayed transaction amount graph to represent the inter-account transaction Bitcoin amount. Both the reachability patterns and transaction amount patterns play an important role in the Bitcoin transaction forecasting task. The second stage of DLForecast development is to utilize node embedding~\cite{perozzi2014deepwalk} to map the transaction account relations into a condensed vector space, and build the Bitcoin transaction forecasting system by training a neural network with the extracted transaction account vectors. The goal is to link the current transaction pattern (in the form of embedding) between two accounts to the probability of the transaction. The dynamics of bitcoin transactions make it challenging to build a once-for-all transaction predictor due to the changing transaction pattern and the short life span of bitcoin transaction accounts. We set up a time slot for the transaction prediction model update. At the beginning of each time slot, we fine-tune the trained forecasting system with transactions and accounts in the previous time slot. By promoting such an on-the-fly evolution of the forecasting model, we provide a reasonably high forecasting accuracy. The third and final stage of the DLForecast development is to combine multiple transaction pattern graphs constructed using different types of extracted features. Due to the changing dynamics of Bitcoin transactions, neither the time-decayed reachability graph nor the time-decayed transaction amount graph is capable of capturing different transaction patterns alone.
Namely, no single feature graph can outperform all others.
This motivates us to develop mechanisms that can combine different graphs constructed from different sets of the extracted features.

To the best of our knowledge, this is the first paper applying DNN models on forecasting Bitcoin transactions using the real-world Bitcoin transaction data. In summary, the paper makes three contributions. First, we capture the transaction reachability of user accounts and Bitcoin transaction amount patterns to provide a unique understanding of the spatiotemporal dynamics of Bitcoin transactions. Second, we develop DLForecast, a Bitcoin transaction forecasting system. The proposed system evolves on-the-fly and is capable of predicting how likely the two accounts will make transactions in the near future. Third but not the last, we apply the Multiplicative Model Updates (MMU) ensemble to combine prediction models trained over different transaction features extracted from the bitcoin transaction graph. The ensemble ensures the stable yet competitive performance of the proposed Bitcoin transaction forecasting system.  We achieve accuracy of over 60\% on the future transaction forecasting and improve the performance by more than 50\% when compared to the forecast model built on the static graph baseline.

The rest of the paper is organized as follows. Section 2 provides the related work. Section 3 presents a statistic analysis on the Bitcoin transaction dataset and Section 4 discusses the design and evaluation of the Bitcoin transaction forecast system. We show the performance improvement of the Bitcoin transaction forecast with the MMU ensemble in Section 5 and conclude the paper in Section 6.

\section{Related Work}

The DLForecast development is inspired by two orthogonal research threads: (1) Statistic characterization of the Bitcoin transaction dataset. (2) Graph Mining.

{\bf Statistical characterization of Bitcoin transaction data.\/} Most of the existing work on the statistical analysis of Bitcoin transaction data falls into this category. \cite{ron2013quantitative} analyzed Bitcoin transactions carried out until May 2012 and discovers that a massive number of transactions only involve a small number of Bitcoins and only a few transactions move a large amount of money. \cite{lischke2016analyzing} analyzed the transaction graph until May 2013, identified an initial phase of growth of the Bitcoin network, and measured network characteristics, temporal patterns, and the wealth accumulation over time. \cite{maesa2016uncovering} studied Bitcoin transaction user graph until December 2015, analyzed the time evolution of Bitcoin network, and verified the \textit{rich get richer} conjecture, i.e., a user with higher balance or number of incoming transactions with respect to other users in the network tends to accumulate even higher balance or more incoming transactions over time. \cite{pareja2019evolvegcn} studied the trust and rating of the bitcoin transaction networks, predicted the polarity of each rating, and forecasted whether a user will rate another one in the next time step.  In recent years,
\cite{greaves2015using,akcora2018forecasting,mcnally2018predicting} utilize the Bitcoin transaction graph data to make Bitcoin price prediction. However, none of the existing work, to the best of our knowledge, has developed a DNN-model-based transaction forecasting system. Example predictions include the likelihood of making a transaction between two accounts, or which account is the most likely to conduct a transaction with a given account.

{\bf Graph Mining.\/} The recent progress on representation learning has extended to complex structures, like networks and graphs. Node embedding on static graphs aims to map the structural information pertaining to a node to produce a low-dimensional representation. Various techniques such as random walks~\cite{perozzi2014deepwalk,grover2016node2vec}, matrix factorization~\cite{cao2015grarep}, edge-sampling~\cite{tang2015line}, and structure learning~\cite{wang2016structural} have been explored for graph mining. Alternatively, convolutional neural networks are used to build GCN (Graph Convolutional Networks) and to capture the hidden relations between nodes and edges of a graph ~\cite{hamilton2017inductive,scarselli2008graph,pareja2019evolvegcn,fu2019core,fu2021mimosa,fu2021probabilistic,fu2021differentiable,fu2022hint}. GCN-based embedding and transaction prediction are beyond the scope of this paper and can be considered as future work. Graph embedding can be used for many applications, such as community detection~\cite{wang2017community,zhang2018cosine}, anomaly detection~\cite{yu2018netwalk}, graph clustering~\cite{yang2017graph}, and link prediction~\cite{jiang2016encoding,zhu2016scalable}. However, these approaches can only work with static graphs and fail to use temporal information to handle evolving graphs.
Many real-world graphs, such as social networks, are evolving. For example, new links can form in a citation network (e.g., when new colleagues are hired or joined the project) and old links may disappear (e.g., when colleagues left the project or the organization).
Recently, dynamic network embedding approaches are proposed to study graphs that evolve~\cite{zhu2016scalable,zhou2018dynamic,nguyen2018continuous,goyal2018dyngem,zuo2018embedding}. However, many existing representation learning techniques for dynamic graphs assume that graph dynamics evolve at a single time scale process. \cite{trivedi2018dyrep} considers two distinct dynamic processes: topological evolution and node interaction evolution at different time scales. Existing dynamic graph techniques can be categorized into two approaches: discrete-time approach and continuous-time approach. The former approach observes the evolution of a dynamic graph as a collection of static graph snapshots over time~\cite{zhou2018dynamic} and the latter models the dynamic graph at a finer time granularity.

Given that the Bitcoin transaction graph is highly dynamic with continuously incoming transactions and new accounts, we propose to leverage dynamic node embedding techniques to explore the hidden transaction patterns in the Bitcoin transaction graph and to forecast future transactions between accounts. To incorporate richer transaction dynamics, we consider features that are intrinsic in the Bitcoin Transaction: short yet diverse length of the user accounts life span and local transaction pattern that only appears in a short period. Motivated by \cite{beres2018temporal}, we represent the dynamic graphs as a collection of snapshots, apply static embedding algorithms to each snapshot, and update the resulting static embedding across time steps.

Unlike many existing graph embedding approaches considering only a single timescale or a single feature, \cite{perozzi2017don,chen2018harp, lei2019grahies} inject hierarchical or multi-scale feature extraction to learn a better representation of the graph. These features are either focused only on the (spatial) graph scale or on the (temporal) time-changing scale. Different from these papers, we combine different spatial and temporal features to capture the dynamics in Bitcoin transactions.  Due to the high dynamics of the Bitcoin transaction and the changing transaction pattern,  which embedding feature has the best ability to capture transaction pattern varies over time.
In a dynamic environment, we iteratively choose transaction forecasting models constructed from embedding from different Bitcoin transaction features without knowledge of the future. A cost(correct or incorrect forecasting) would be paid based on the forecasting decision and the observed outcome. In both game theory and machine learning literature, a host of algorithms are proposed to make decisions that are nearly as well as the best single decision in hindsight~\cite{freund1997using,herbster1998tracking,cesa2006prediction}. While most of these works are based on the assumption of a fixed outcome distribution, e.g. the transaction pattern of the accounts does not change over time and therefore multiple fixed prediction models can be used to explore different patterns as each model is an expert in predicting a certain type of node relations(sparse or dense, for example). However, the Bitcoin transaction graph is highly dynamic, and transaction pattern changes over time. In this case, the underlying outcome distribution changes. For example, nodes with sparse connections tend to have more transactions in the past and may tend to stay inactive recently. Consequently, a good forecasting model for such nodes in the past may not be effective now due to the changing transaction behavior of the node. Therefore, it is inappropriate to keep a fixed set of forecasting models. Online portfolio management algorithms should be applied to keep a dynamic choice of the forecasting models in the changing environment~\cite{hazan2009efficient,li2014online,besbes2014stochastic}.

\section{Bitcoin Dataset and its Statistic Analysis}

We first provide an introduction to the real-world Bitcoin transaction dataset and demonstrate three key features: reachability pattern, transaction amount pattern, and dynamics.

\subsection{Introduction to the Bitcoin transaction dataset}

We consider a Bitcoin transaction dataset~\cite{kondor2014rich} containing
298,325,122 Bitcoin transactions in the first 508241 blocks, i.e. from Jan $3^{\text{rd}}$ 2009 to Feb $9^{\text{th}}$, 2018. There are four fields in the data format:
$$\left\langle \text{txid}, \text{in\_addr}, \text{out\_addr}, \text{weight} \right\rangle$$
\textbf{Txid} is the index of the transaction. Within one \textbf{txid},
a transaction with inputs from
$m$ distinct sender addresses (\textbf{in\_addr}) and outputs to $n$ distinct receiver addresses (\textbf{out\_addr}) is processed to $m \times n$ directed edges. While one address can be considered as one account and one user can have multiple accounts for transactions, there are 297,816,881 unique accounts in 298,325,122 transactions.
The edges are weighted according to the Bitcoin values transferred between accounts.
Note that addresses that could not be decoded in the aforementioned dataset are labeled with a special address value of $-1$. The number of Bitcoins transferred is written in Satoshis, i.e., $10^{-8}$ Bitcoin.
Note that the dataset does not include any information on transaction fees nor mining transactions (transactions with zero inputs). For Transaction forecasting, the transaction fee and the mining reward should be processed separately.
We provide the statistics in Table~\ref{table:stat_all}.

\begin{table}[t]
\centering
\scalebox{1}{
\small{
\begin{tabular}{|l|r|}
\hline
\# blocks & 508,241 \\ \hline
\# accounts & 297,816,881 \\ \hline
\# transactions & 298,325,122 \\ \hline
\# sender-receiver pairs & 2,536,261,805 \\ \hline
\end{tabular}
}}
\caption{Statistics of the Bitcoin transaction dataset}
\label{table:stat_all}
\vspace{-0.3cm}
\end{table}

We make two key observations that are essential to the subsequent analysis and task of transaction forecasting between accounts. (1) Since each transaction involves $m$ senders and $n$ receivers, one \textbf{txid} involves multiple sender-receiver pairs. In total, there are 2,536,261,805
sender-receiver pairs in 298,325,122 transactions. When defining new addresses as those that are not in the existing graph and old addresses as those that are already in the graph, we observe that 60.62\% of the pairs are old addresses sending to other old addresses. 39\% of the pairs are old addresses sending Bitcoins to new addresses. 0.263\% of the pairs are new addresses sending to old addresses. And 0.104\% are new addresses sending to new addresses. (2) An address is designed to be a single-use token, meaning that the address is used in only one transaction. However, people do not change their transaction address as frequently so that some account IDs would appear in multiple transactions. The life span of the Bitcoin address allows us to forecast transactions in a limited period.
As no existing graph mining method can handle the complex transactions between $m$-senders and $n$-receivers, we will use sender-receiver pairs for our study instead of transactions. In other words, we will forecast sender-receiver pairs in future transactions using the features extracted from sender-receiver pairs in existing transactions.

By the anonymity design of the Blockchain, the identity of Bitcoin users cannot be verified unless we have external ground truth information from the real world.
Based on the assumption that addresses that appeared together in a single transaction can be considered as from one user, \cite{greaves2015using} applies the Union-Find algorithm to link addresses that are expected to belong to the same user. However, the assumption would depreciate as indicated in~\cite{ron2013quantitative}: It would either suffer from underestimation in which different addresses that belong to the same user do not necessarily appear in the same transaction or overestimation in which addresses within one transaction do not necessarily belong to the same user. Besides, some newly proposed chain~\cite{kumar2017traceability} even purposely obfuscate transactions from a single entity. Since \cite{ron2013quantitative} shows that statistics in the contracted entity graph is very similar to the original address graph, we do not verify the ownership of the addresses but only use the address graph to forecast transactions.

\subsection{Reachability, Dynamics, and Transaction Pattern}

We present three key features of the Bitcoin transaction data: reachability, dynamics, and transaction amount pattern. Reachability describes the topological connectivity pattern of Bitcoin accounts on how different accounts do transactions, or how different nodes are connected in the Bitcoin transaction graph. For example, two accounts that never make transactions with each other will not be connected with an edge in the graph. Transaction amount pattern shows how much Bitcoin is sent or received in a transaction and it can be considered as the weight attribute of the reachability edge. The dynamics of the Bitcoin transaction data indicates the frequency of the transaction and reflect the activeness and the life duration of the Bitcoin accounts. While these features are not unique for the Bitcoin transaction data but to all dynamic graphs, they reveal the transaction behavior of Bitcoin users.

\begin{table}[t]
\centering
\scalebox{0.72}{
\small{
\begin{tabular}{|c|c|c|c|c|c|c|}
\hline
\multirow{2}{*}{} & \multicolumn{3}{c|}{first 100k} & \multicolumn{3}{c|}{latest 100k} \\ \cline{2-7}
 & \#pairs & \#transactions & \#accounts & \#pairs & \#transactions & \#accounts \\ \hline
time1 & 10006 & 560 & 8247 & 10013 & 719 & 6153 \\ \hline
time2 & 20013 & 4232 & 14396 & 20310 & 1546 & 9659 \\ \hline
time3 & 30001 & 7756 & 23658 & 30004 & 2010 & 11558 \\ \hline
time4 & 40001 & 12505 & 31907 & 40001 & 3503 & 17072 \\ \hline
time5 & 50002 & 17446 & 37729 & 53069 & 4228 & 21198 \\ \hline
time6 & 60002 & 22195 & 44479 & 60077 & 5858 & 26454 \\ \hline
time7 & 70001 & 27508 & 49187 & 70005 & 6708 & 32401 \\ \hline
time8 & 80001 & 31533 & 56430 & 83246 & 8418 & 38138 \\ \hline
time9 & 90323 & 35422 & 63065 & 90090 & 9584 & 41699 \\ \hline
time10 & 100824 & 39010 & 69971 & 100045 & 10013 & 50441 \\ \hline
\end{tabular}
}}
\caption{Reachability of the Bitcoin transaction subsets}
\label{table:stat}
\vspace{-0.3cm}
\end{table}

\begin{figure}[t]
\begin{minipage}{0.49\linewidth}
 \centerline{\includegraphics[scale=.24]{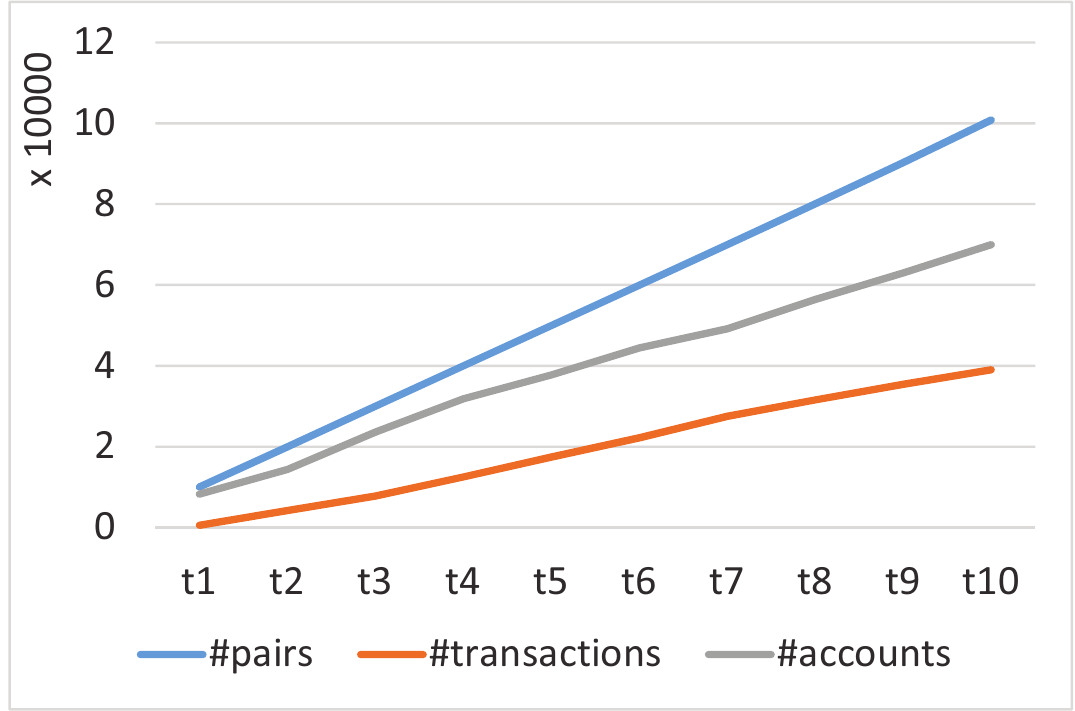}}
 \subcaption{First 100k}
\end{minipage}
\begin{minipage}{0.49\linewidth}
 \centerline{\includegraphics[scale=.24]{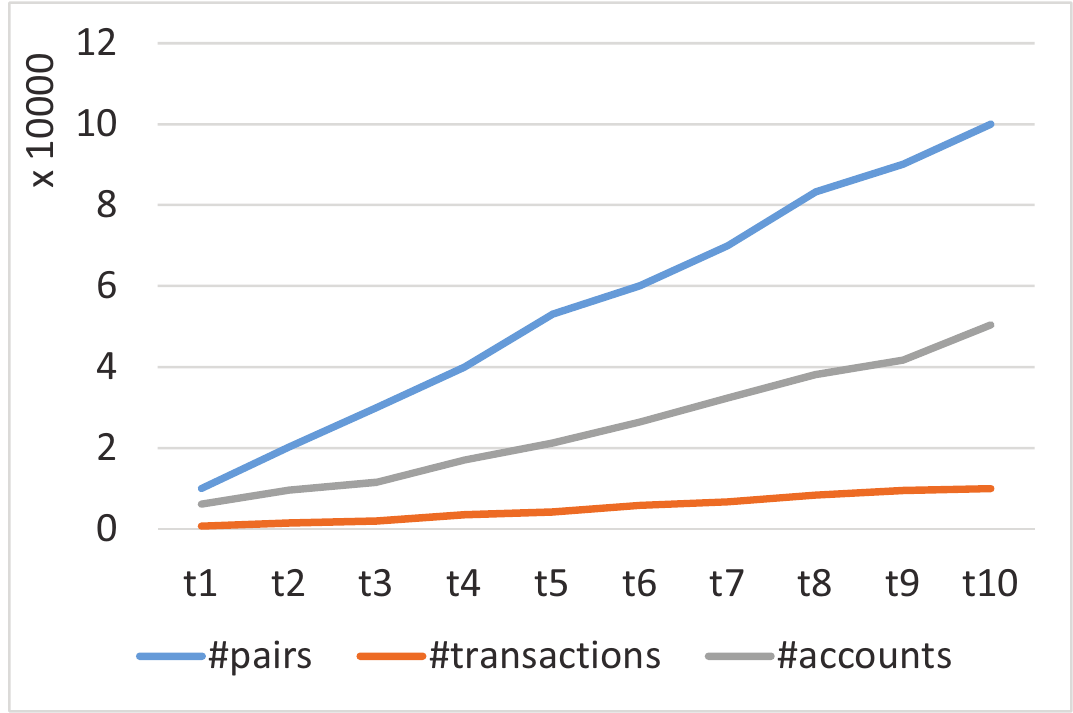}}
  \subcaption{Latest 100k}
\end{minipage}
\caption{Increase of \#accounts and \#transactions in 100k sender-receiver pairs from inception of Bitcoin and the latest of the dataset}
\label{figure:stat}
\vspace{-0.3cm}
\end{figure}
%Figure~\ref{figure:stat}

For ease of representation and analysis,
we consider two representative subsets of the full Bitcoin transaction data: the first 100824 sender-receiver pairs from 39,010 transactions in 38,708 blocks at the beginning of the Bitcoin launch (from Jan $3^{\text{rd}}$, 2009 to Feb $6^{\text{th}}$, 2010) and the latest 100045 sender-receiver pairs from 10,013 transactions in 6 blocks at the end of the dataset (from 9:37 am to 10:56 am on Feb $8^{\text{th}}$, 2018).
The two subsets are sampled from completely different periods of time and demonstrate very different node reachability patterns. With approximately 10k sender-receiver pairs as the interval, we provide the statistics of sender-receiver pairs, transactions, and accounts in Table~\ref{table:stat}. The increasing popularity of the Bitcoin has increased both the average number of sender-receiver pairs in a transaction and the total number of transactions per block. To be specific, it takes 39010 transactions and 38708 blocks to have 100k sender-receiver pairs at the inception of Bitcoin and it takes 10013 transactions and 6 blocks at the end of the provided dataset. Meanwhile, the total number of 50441 accounts in 10013 transactions at the end of the dataset is much denser than the 69971 accounts in 39,010 transactions. We visualize the trend in Figure~\ref{figure:stat}.
The complex relationship between $m$-senders and $n$-receivers at the end of the dataset makes the latest 100k sender-receivers more difficult to process than the first 100k pairs.
Note that partitioning timestamp using the number of sender-receiver pairs is just one way. Other time-series information can also be explored, such as partition using actual time, e.g. by the hour, the day, or the week, and using the number of transactions, e.g. every 100 transactions.

\begin{figure}[t]
\begin{minipage}{0.49\linewidth}
 \centerline{\includegraphics[scale=.34]{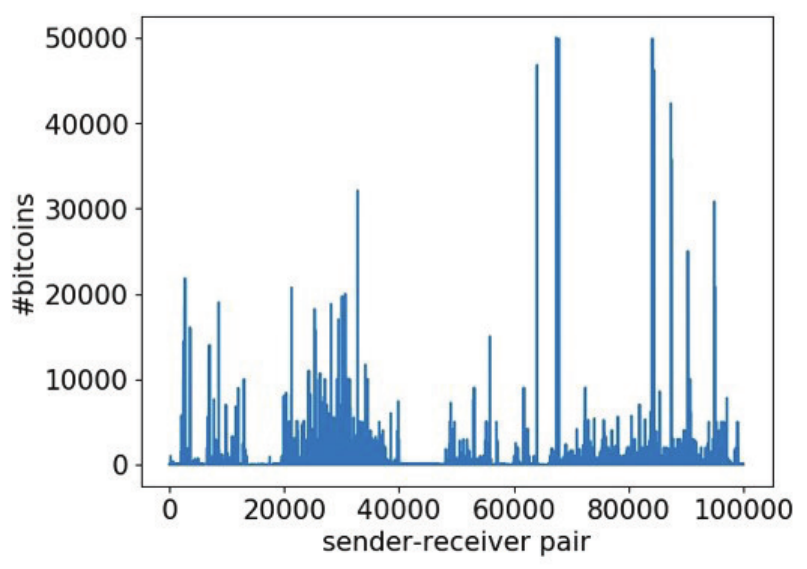}}
 \subcaption{First 100k}
 \label{figure:amount_first}
\end{minipage}
\begin{minipage}{0.49\linewidth}
 \centerline{\includegraphics[scale=.34]{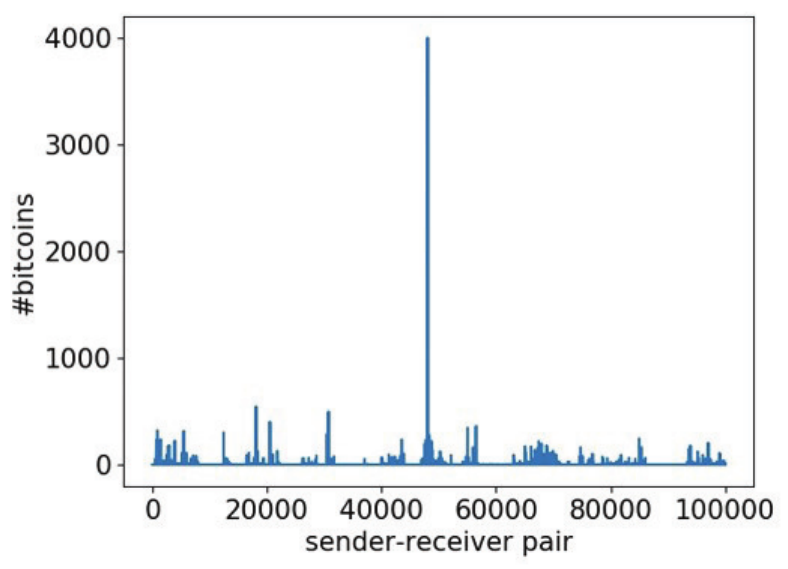}}
  \subcaption{Latest 100k}
   \label{figure:amount_last}
\end{minipage}
\caption{Bitcoin transaction amount pattern: \#Bitcoins per sender-receiver pair}
\label{figure:bitcoin_amount}
\vspace{-0.3cm}
\end{figure}

\begin{figure}[t]
 \centerline{\includegraphics[scale=.49]{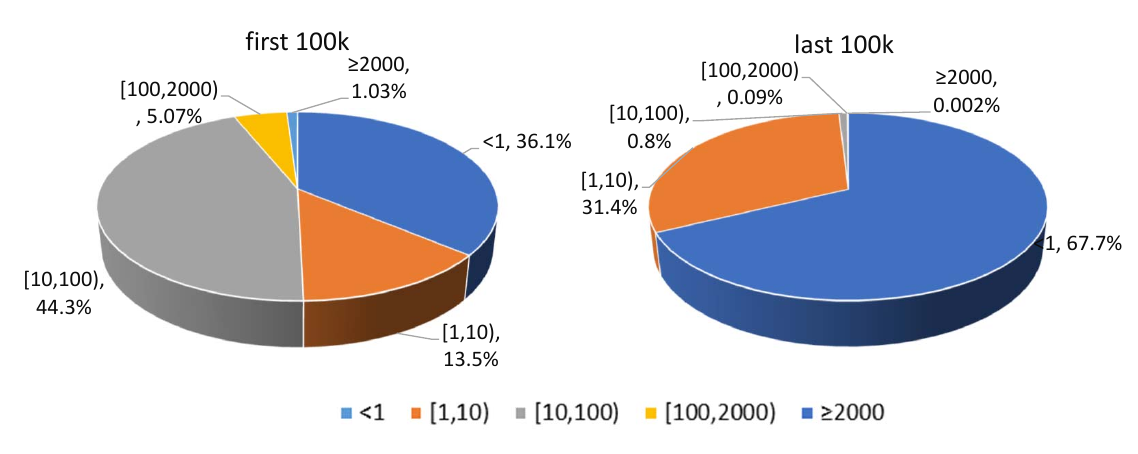}}
\caption{Bitcoin transaction amount frequency in the sender-receiver subset: first 100k sender-receiver pairs and the latest sender-receiver pairs}
\label{figure:reachability}
\vspace{-0.3cm}
\end{figure}

The transaction amount pattern is another interesting feature of the Bitcoin transaction. In Figure~\ref{figure:bitcoin_amount}, we show that (1) the number of Bitcoins transferred between accounts changes over time; (2) there are some local features in the Bitcoin transaction amount, e.g. low transaction amount during the first 40,000 to 50,000 sender-receiver pairs. (3) transactions with an extremely large number of Bitcoins are rare. The average transaction amount of 93.7 Bitcoins in the first 100k sender-receiver pairs is higher than the averaging 0.58 Bitcoins in the latest 100k, showing some difference in transaction amount pattern in the two subsets. (4) the Bitcoin amount in most transactions would fall into some space, i.e. $\le100$ Bitcoins in the first 100k sender-receiver pairs and $\le10$ Bitcoins in the latest 100k sender-receiver pairs. We demonstrate the distribution of the Bitcoin amount in Figure~\ref{figure:reachability}.

\begin{figure}[t]
 \centerline{\includegraphics[scale=.55]{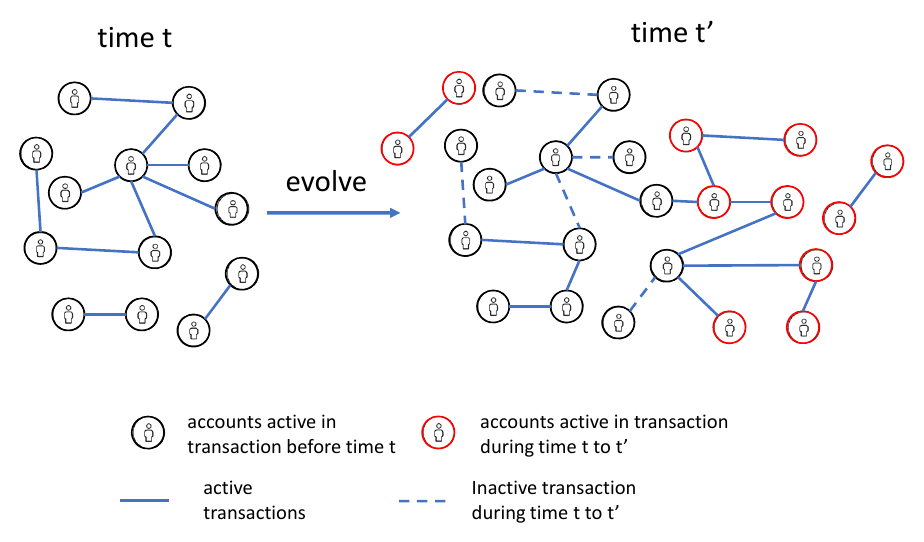}}
\caption{Illustration of Bitcoin transaction dynamics: new addresses and new transactions emerge while previous  transaction addresses become inactive}
\label{figure:dynamics}
\vspace{-0.3cm}
\end{figure}

Besides the reachability and transaction pattern features,
dynamics is another key feature of the Bitcoin transaction. When each address is considered as one node, and each sender-receiver pair represents an edge in the graph, we illustrate the evolving dynamics of the Bitcoin transaction in Figure~\ref{figure:dynamics}. During time $t$ to time $t'$, which is some time later than time $t$, new transactions with new accounts are injected to the graph and some previous addresses and transactions become inactive.
If the life span of the address runs out, the nodes will no longer be involved in any transactions and can be deleted.
We observe that the length of the life span depends on the frequency of transactions. In particular, the life span of a Bitcoin address is longer at the inception period than the life span right now as transactions are more frequent today.

\section{Dynamic Bitcoin Transaction Forecasting}

Due to the highly dynamic transaction pattern of the bitcoin transactions, it is challenging to leverages these dynamics while exploiting the historical transaction data for future transaction forecasting.
In this section, we first elaborate the construction of the time-decaying reachability graph and the time-decaying transaction amount graph from the Bitcoin transaction data while considering the network dynamics. Then, we demonstrate how to perform node embedding on the constructed graphs and how to build the Bitcoin transaction forecasting model using neural networks. Initial experiment results are provided to demonstrate the effectiveness of the proposed forecasting techniques.

\subsection{Spatiotemporal Graph Construction}

A graph $G=\{ V, E \}$ has $N=|V|$ number of vertices and $E$ number of edges. A straightforward way to construct the Bitcoin transaction graph is considering sender addresses and receiver addresses as nodes, sender-receiver pairs as edges, and Bitcoin amount as weight. Since one address may involve in multiple transactions, there can be multiple single-direction edges from the sender vertex to the receiver vertex over time. Besides,
the role of the sender and the receiver can also switch. As no graph mining algorithm can process such complicate repeated, weighted and directed connectivity between nodes, it is natural to simplify the problem.

To extract observable transaction features from the Bitcoin transaction data, we model the Bitcoin transaction data using two types of spatial relations between a pair of accounts. At first, we take advantage of the number of transactions between two accounts and build a reachability graph where the edge weight $w_{t_i}=0$ when there is no connection between two accounts and $w_{t_i}=1$ as long as there is a connection. $t_i$ presents the timestamp. Similarly, we make use of the number of Bitcoins sent between two accounts and build a transaction pattern graph. Edge weight of the transaction pattern graph $w_{t_i}=0$ when there is no Bitcoin sent between two accounts, edge weight $w_{t_i}=1$ when the number of Bitcoins falls into the frequent transaction range (for example, $\le 10$ Bitcoins in the latest 100k sender-receiver pairs), and edge weight $w_{t_i}=0.5$ when the number of Bitcoins falls into the occasional transaction range. In both graphs, the weight is designed to describe the transaction behavior of two nodes. Accordingly, the forecasting task in this paper would focus on the sender-receiver pair between two accounts rather than the transaction between $m$-senders and $n$-receivers. Both simplified graphs are undirected and so the forecasting concerns only on the probability of two accounts that may transact but does not indicate the sender-receiver relationship.

To capture the dynamics in Bitcoin transactions, we further incorporate temporal evolving information between a pair of accounts by a time-decay factor $\alpha$. Assuming the time period of the data collection is divided into $m$ periods, we use $\alpha_{t-\Delta t}=\frac{1}{2^{\Delta t}}$ in our first Bitcoin transaction forecasting prototype. The opt-out threshold is a hyperparameter, which is tunable over time.
Opt-out means that an account in the form of node is deleted from the graph due to its inactivity or its overall short lifespan of the account in Bitcoin transactions.
The opt-out threshold of 0.125 is empirically chosen given the dynamic of the Bitcoin transactions, indicating that if there is no new transaction for a given account in consecutively 3 periods, we will delete the account from the graph due to the limited lifespan of the accounts in Bitcoin transaction. In short, the edge weight on the constructed time-decayed reachability graph and the time-decayed transaction pattern graph is formulated as \begin{equation}
    W_t({u,v}) = \alpha_{t - t_0}w_{t_0} + \alpha_{t - t_1}w_{t_1} + ... + \alpha_{t - t_m}w_{t_m}+w_t.
    \label{equa:weight}
\end{equation}
Note that this threshold is set to accommodate the dynamic transaction pattern of the Bitcoin transactions and it does not necessarily to be fixed.
We also take a static graph as the baseline. The static graph only considers the topology of the transaction data and only capture if there is a transaction between two accounts but not how often or how much amount. For all three graphs, we forecast that given two accounts (addresses) at time $t$, how likely they are to trade in the near future, namely from time $t$ to time $t+t_m$.

While we represent the dynamic graph as a collection of snapshots on whole-data, stratified random sampling of the original data can work with a much smaller dataset and provide forecasting. However, since the Bitcoin transaction is highly dynamic and the lifespan of a single transaction address varies, multiple time-decay factors are used to learn the transaction patterns on-the-fly. With whole-data, we dynamically evaluate the impact of recent transactions and past transactions at each time-step and train the prediction model accordingly. Although applying stratified random-sampling directly may not capture such a dynamic transaction pattern, it can be another way of investigating the Bitcoin transaction data.

\subsection{Node Embedding in Dynamic Transaction Graph}

A primary tool to analyze Bitcoin transaction relations is the $N \times N$ adjacency matrix, in which $N$ is the number of accounts in the graph. Each column and each row in the matrix present a node. Non-zero values in the matrix indicate that two nodes are connected. While many graph mining algorithms fit the entire adjacency matrix in memory, it is intractable when there are a large number of nodes in the graph. To scale the processing of large-scale Bitcoin transaction graph, we seek to use more compressed representation with richer features beyond the sparse adjacency matrix.
We appeal to graph embedding, which maps the node relations into a much more condensed format using a vector space model.

\begin{figure}[t]
\begin{minipage}{0.49\linewidth}
 \centerline{\includegraphics[scale=.29]{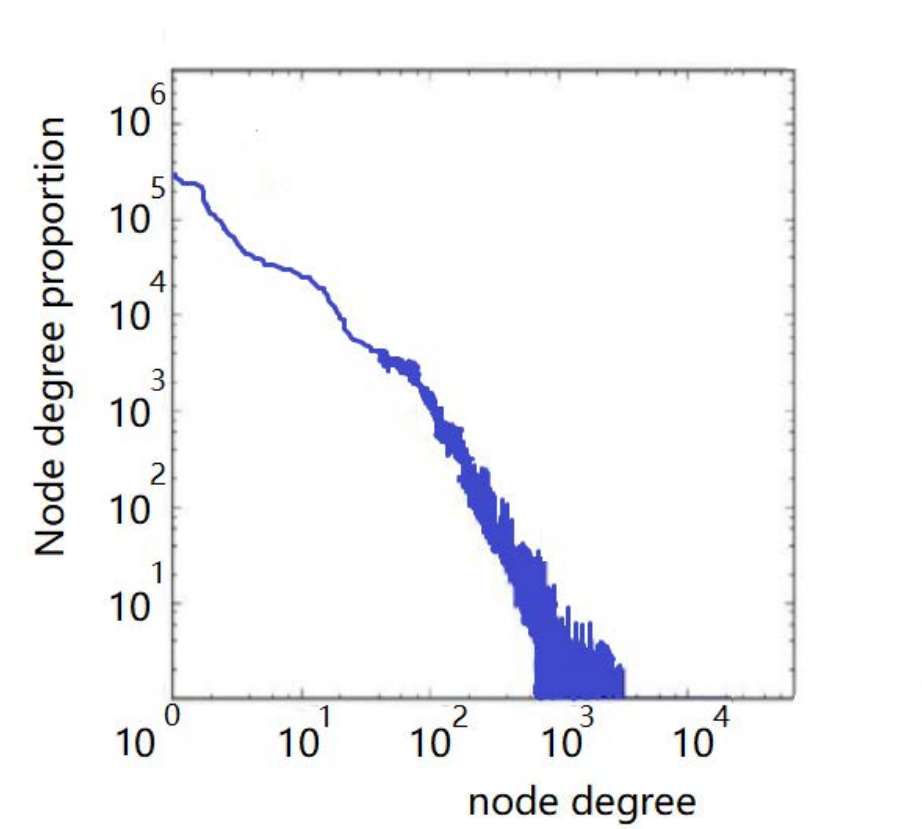}}
 \subcaption{Indegree distribution}
 \label{figure:in_degree}
\end{minipage}
\begin{minipage}{0.49\linewidth}
 \centerline{\includegraphics[scale=.29]{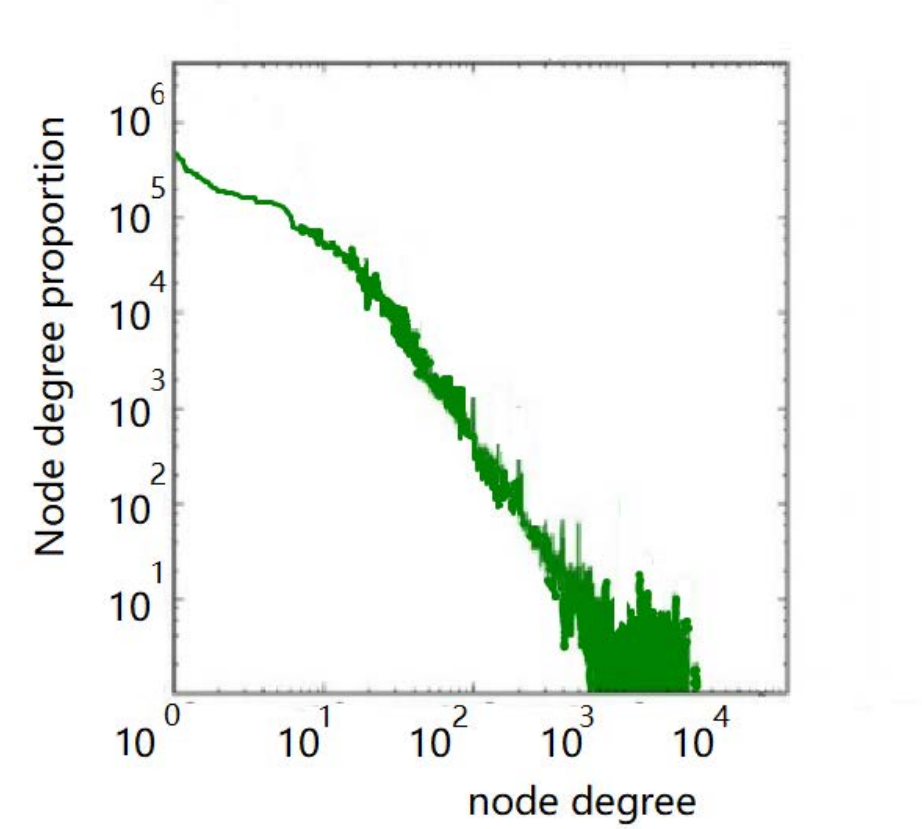}}
  \subcaption{Outdegree distribution}
   \label{figure:out_degree}
\end{minipage}
\caption{Transaction pattern power-law: indegree and outdegree for the first 100k sender-receiver pairs}
\label{figure:power_law}
\vspace{-0.3cm}
\end{figure}

\begin{figure}[t]
 \centerline{\includegraphics[scale=.50]{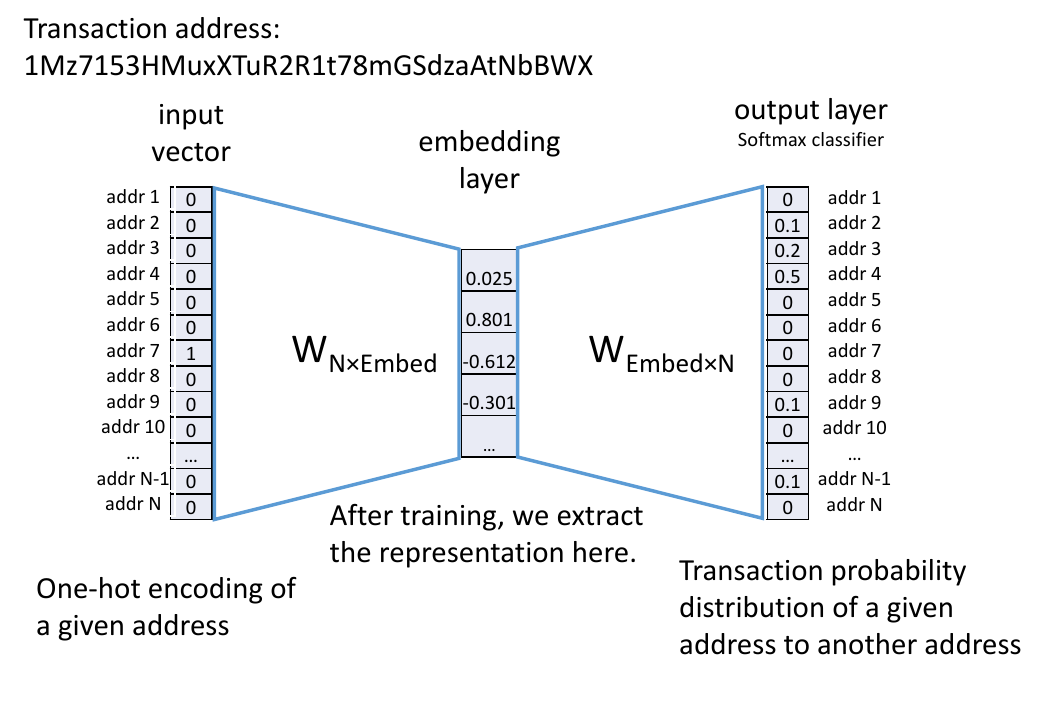}}
\caption{Skip-Gram deep learning model architecture}
\label{figure:skipgram}
\vspace{-0.3cm}
\end{figure}

\begin{figure*}[ht]
 \centerline{\includegraphics[scale=.62]{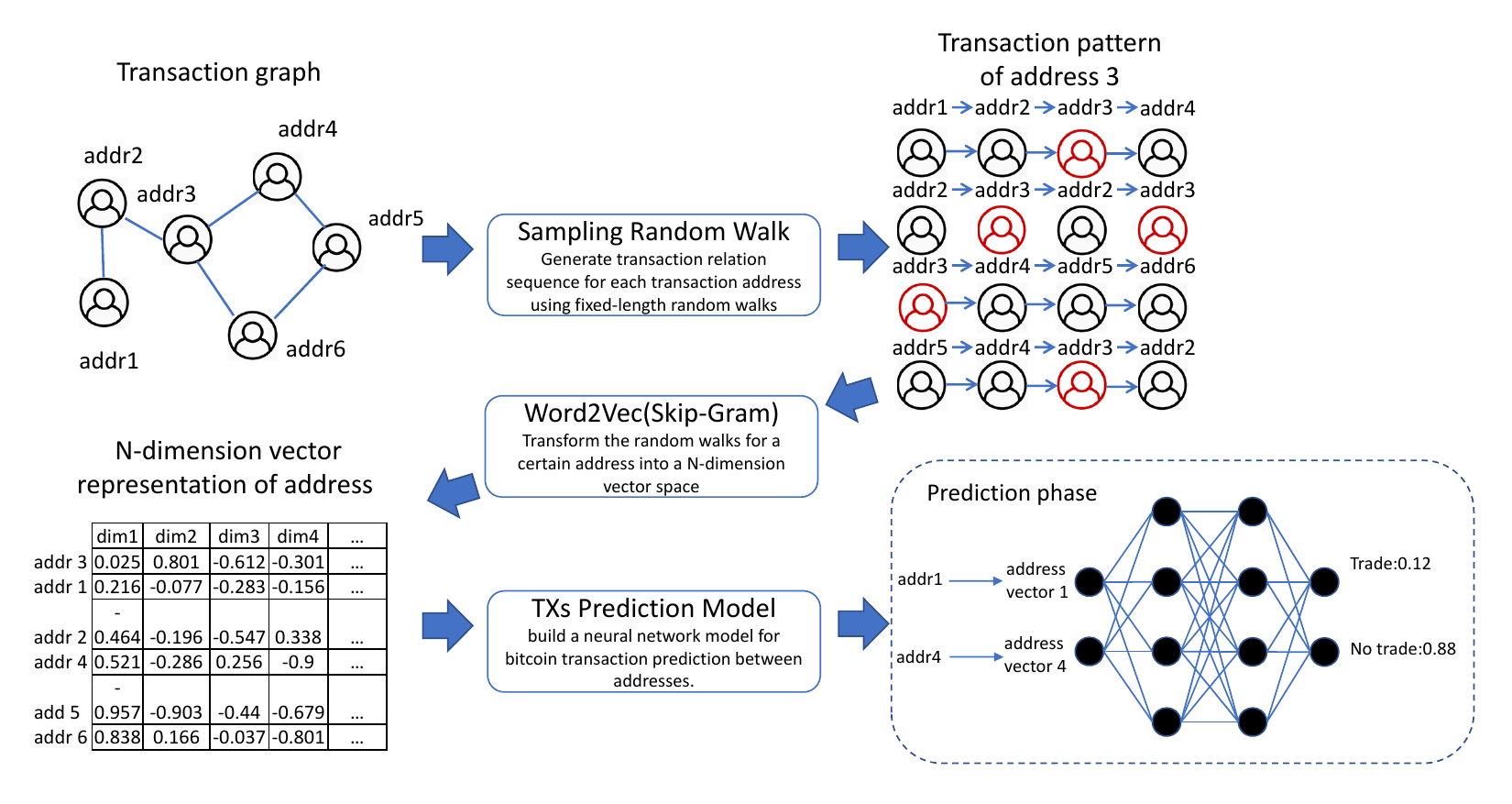}}
\caption{The training workflow of Bitcoin transaction forecast. At the beginning of each time slot, the TXs prediction model is onlinely updated using the transaction data from the previous slot.}
\label{figure:workflow}
\vspace{-0.3cm}
\end{figure*}

The idea of putting graph data into compressed embedding is inspired by the fact that
the indegree and outdegree of the Bitcoin accounts in the transaction graph follow the power-law distribution as shown in Figure~\ref{figure:power_law}. The power law indicates that most of the Bitcoins are held by a few accounts while most of the accounts last briefly and make transactions with a very small amount of the Bitcoin. Similar to the word frequency in natural language, which also follows the power-law distribution that there are only a few words that are frequently used, the task of forecasting if two nodes are more likely to have a transaction can be modeled as finding two words that are prone to co-appear.
The short random walks for a specific node on the graph can be modeled as sentences containing a specific word. Since words that are semantically similar are used in similar contexts and these embedding encode the semantic meaning of words such that semantically similar words will lie close to each other in that vector’s space, accounts that make transactions more often would have a closer representation in low-dimensional vector space.

There are 2 steps in node embedding: random walk and word2vec. Similar to \cite{beres2018temporal}, we run a temporal random walk algorithm as step 1. When new edges $(u,v)$ arrive at timestamp $t$, we update all walks ending at node $u$ with a decay factor as described in equation~\ref{equa:weight}. For computational efficiency, walks are deleted if their time-decayed weight becomes very small, e.g., the threshold of 0.125 as indicated in previous sections. We generate 10 randoms walks for each account, to build the context of that account and each random walk has a multi-hop length of 40. Note that the performance of DLForecast is dependent of the choice of these parameters. When the number of random walks is too small and the length of the walk is too short, the generated embedding may have an incomplete and biased representation of the node relation.  We take the embedding parameter setting from~\cite{perozzi2014deepwalk} due to the similar scale of social networks and Bitcoin transaction networks.
%to fully explore the connectivity of Bitcoin transaction accounts and to prevent large computation overhead.

In step 2, the Skip-Gram algorithm is used to map the one-hot encoded representation of the node in the graph to the hidden embedding space. As illustrated in Figure~\ref{figure:skipgram}, Skip-Gram is performed using a neural network model with one hidden layer. The input vector is represented as a one-hot vector with $N$ components, one for each account in the account list. A ``1'' is in the position corresponding to a given address (\textit{addr 7} in the example), and 0s are in all of the other positions. The output of the network is also a single vector with $N$ components, indicating the transaction probability distribution of all addresses given an address. The embedding size $E$ equals the dimension of the hidden layer. We choose the embedding dimension of 128 empirically due to the similar size of the social networks and the Bitcoin transaction graph. The network is trained on address pairs sampled from the random walks: \{target address, context address\}. During training, the input is a one-hot vector representing the target address and the output is a one-hot vector representing the context address. When evaluating, the output vector will be a probability distribution of all possible transaction addresses given an address. While constructing the Bitcoin transaction forecasting model, we take the embedding representation in the hidden layer to represent a given address.

\subsection{Constructing Transaction Forecasting Model}

\begin{figure*}[ht]
\begin{minipage}{0.49\linewidth}
 \centerline{\includegraphics[scale=.33]{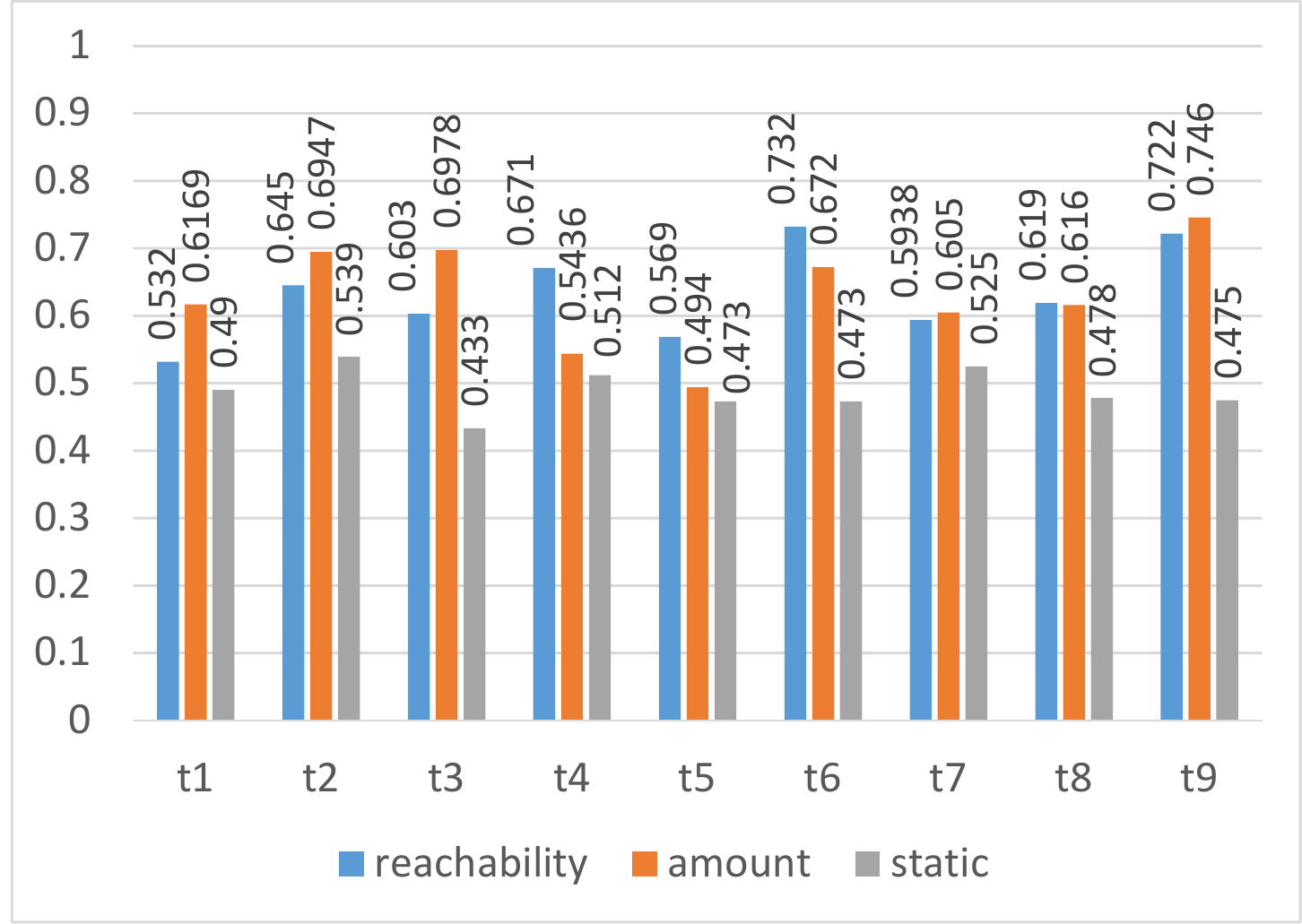}}
 \subcaption{Accuracy first 100k}
 \vspace{-0.3cm}
 \label{figure:acc_first}
\end{minipage}
\begin{minipage}{0.49\linewidth}
 \centerline{\includegraphics[scale=.33]{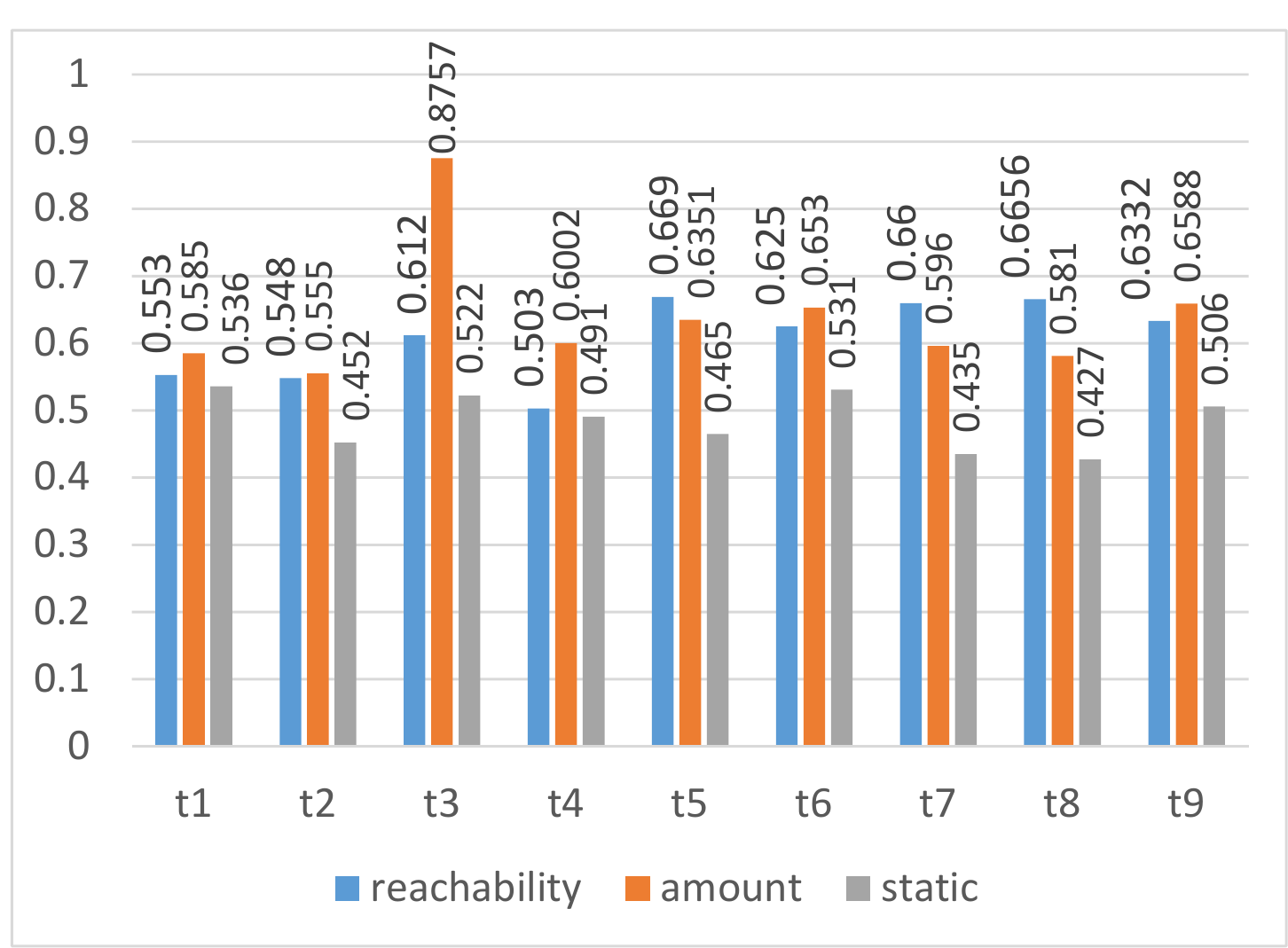}}
  \subcaption{Accuracy latest 100k}
   \label{figure:acc_last}
\end{minipage}
\caption{Transaction forecasting accuracy}
\label{figure:experiment_acc}
\vspace{-0.3cm}
\end{figure*}

\begin{figure*}[ht]
\begin{minipage}{0.49\linewidth}
 \centerline{\includegraphics[scale=.33]{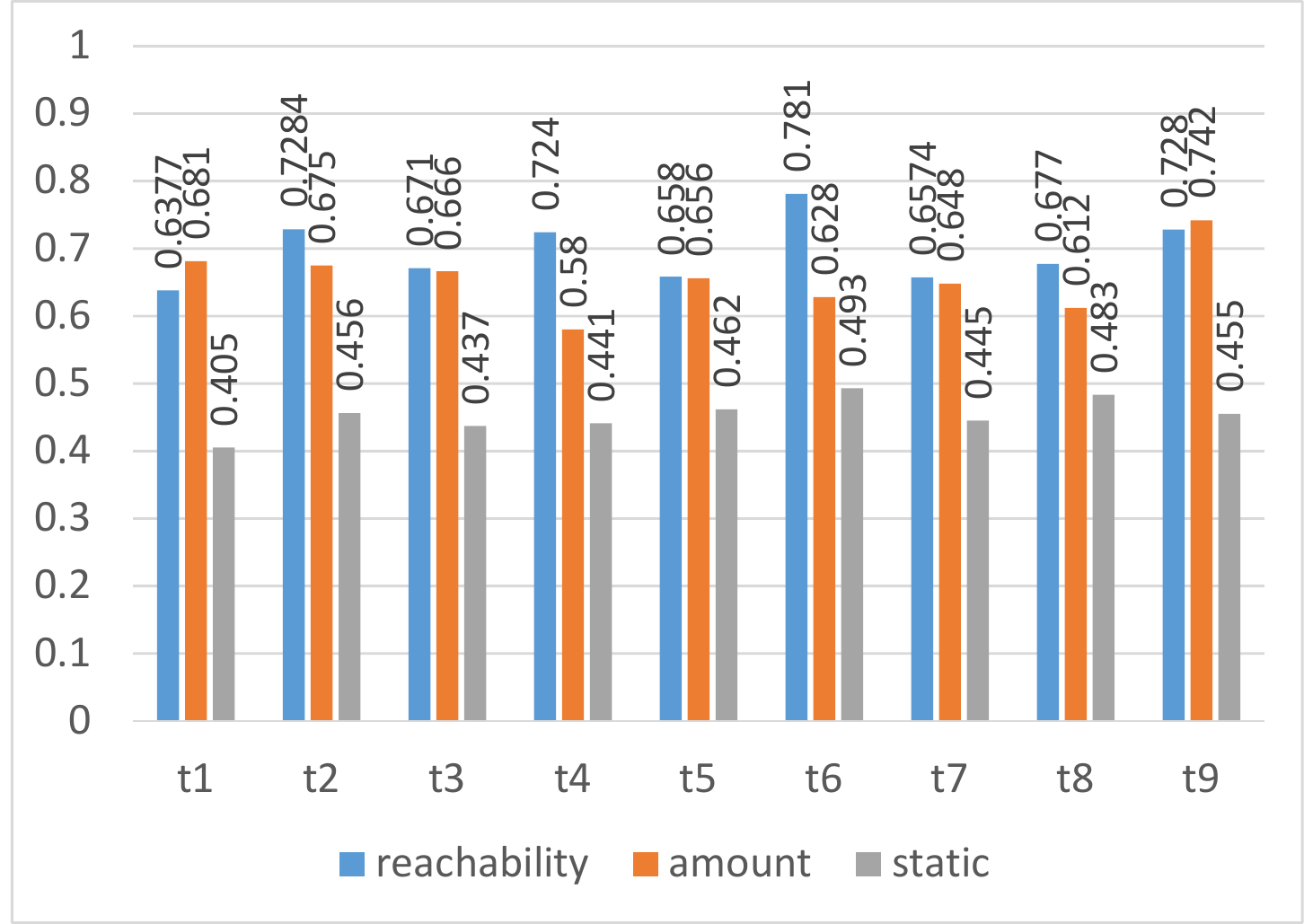}}
 \subcaption{F1 score first 100k}
 \vspace{-0.3cm}
 \label{figure:f1_first}
\end{minipage}
\begin{minipage}{0.49\linewidth}
 \centerline{\includegraphics[scale=.33]{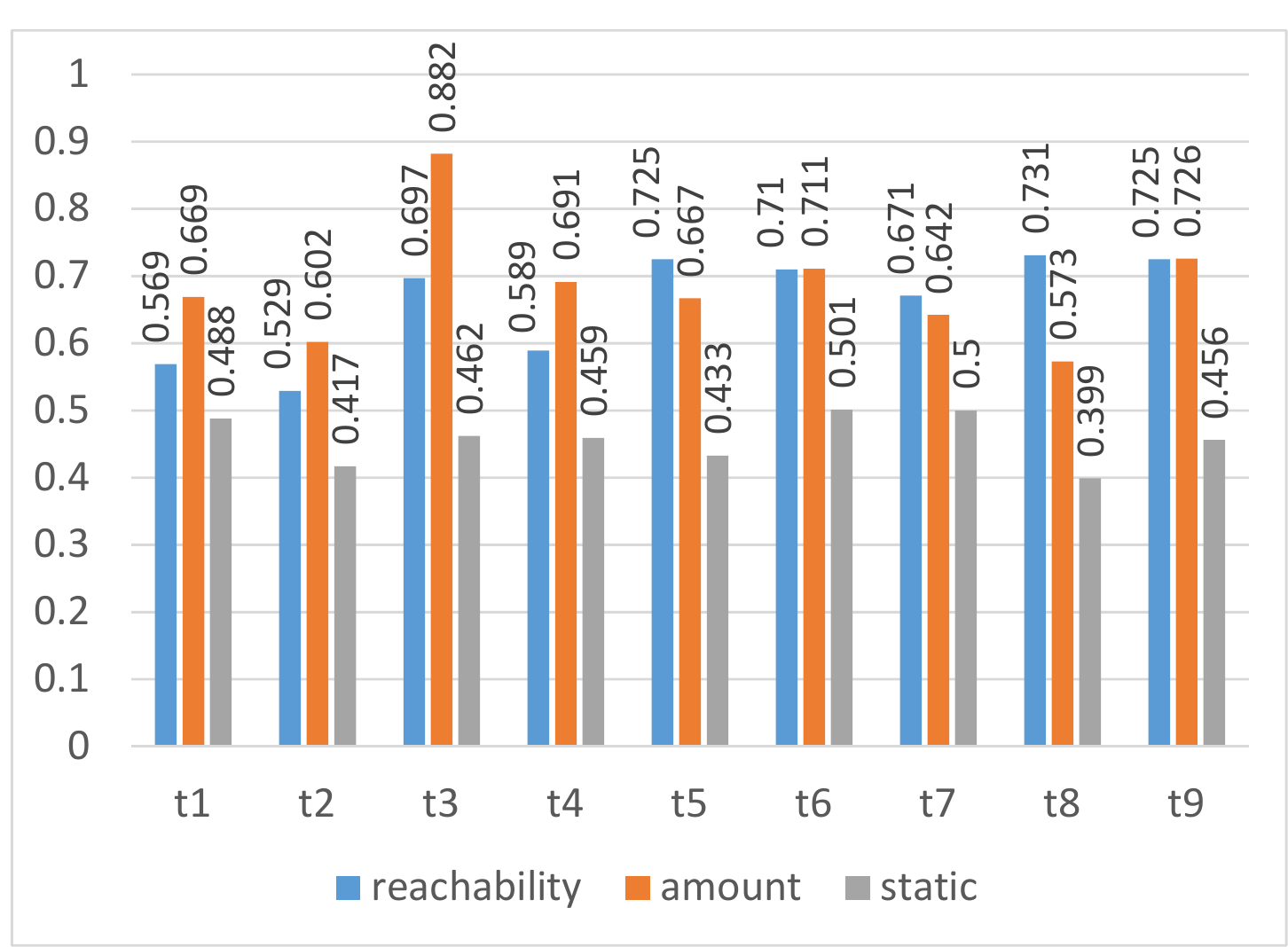}}
  \subcaption{F1 score latest 100k}
   \label{figure:f1_last}
\end{minipage}
\caption{F1 score of transaction forecasting }
\label{figure:experiment_f1}
\vspace{-0.3cm}
\end{figure*}

%Since historical transaction pattern information has been encoded in the embedding $(h_u, h_v)$, we base the forecasting on training a neural network model that links the current transaction pattern with the probability of transaction in the near future.

The deep forecasting model is formed by several successive layers of neurons from the input data to the output. Each layer can be formulated as $f_i=h_i(\mathbf{W}_i x+\mathbf{b}_i)$, where $W$ and $\mathbf{b}_i$ indicate the weight matrix and bias vector, $i \ge 1$ denotes the $i^{th}$ layer and $h_i(\cdot)$ denotes the non-linear activation function. We use $i=3$ in our prototype. $h()$ is the ReLU function and $X$ is comprised of the concatenation of two 128-dimension embedding for the two accounts.
We leverage the transaction information up to time $t$ and forecasts the existence probability of a transaction between address $u$ and $v$, i.e., the probability of an edge $(u,v)$ appear in the Bitcoin transaction graph, from time $t$ to time $t + t_m$.

The training will not be scalable if we use all existent and nonexistent edges because existent edges are substantially fewer than nonexistent ones. Hence, negative sampling is introduced to balance the number of existent and nonexistent edges in both training and test data. To be specific, let $P_t$ and $\bar P_t$ be the ground truth label for existent and nonexistent edges at time $t$ and let $y^t_{u,v}$ be the probability of a future transaction for the binary label: with or without a transaction. The cross-entropy loss $l^t_{u,v}$ of the forecasting model can be defined as
$$
\forall (u,v) \in P_t, \bar P_t,  l^t_{u,v}=-P_t\log(y^t_{u,v})-\bar P_t \log(1-y^t_{u,v})
$$

We provide the bitcoin transaction forecast procedure in Figure~\ref{figure:workflow}.
We train the forecasting model at the end of each time slot (or at the beginning of a new time slot) when all ground truth transaction labels within the time slot are revealed. Note that unless the starting period, only fine-tuning is performed to accommodate new transaction patterns and there is no need to train the new forecasting model from scratch.

We consider an interval of 10k sender-receiver pairs, meaning that we make node embedding every 10k sender-receiver pairs. Specifically, we use embedding generated from pairs 0-10k (from time 0 to time 1) to train a neural network model for forecasting in time 1 to time 2 (pair 10k-20k). Then at the end of time 2, we generate embedding on pair 10k-20k to fine-tune the neural network and use the new prediction model to forecast sender-receiver pairs from time 2 to time 3. Since there are 10 partitions for each subset of the data, we use t1 to t9 to represent the point-of-time in the temporal partitions of the Bitcoin dataset. T1 is the end of time 1 and t9 is the end of time 9.
We evaluate the forecasting performance of Bitcoin transactions using accuracy and f1-score. Accuracy reported at t1 is trained using the graph in partition time 1 and tested on data in time 2. Accuracy at t9 is
trained on weighted data from time 1-time 9 and tested on data in time 10.

\textbf{accuracy.\/} percentage of both positive samples indicating a transaction between two nodes and negative samples denoting no transaction between two nodes that are correctly predicted. It is formulated as $\frac{tp+tn}{(\#total)}$ where $tp$ is the number of true positives and $tn$ is the number of true negatives.

\textbf{f1 score.\/} the harmonic mean of Precision and Recall: $f1=\frac{2*\text{precision}*\text{recall}}{\text{precision}+\text{recall}}$. Precision is the ratio $\frac{tp}{(tp+fp)}$ where $fp$ the number of false positives. Recall is the ratio $\frac{tp}{(tp+fn)}$ where $fn$ the number of false negatives.

We provide the experiment results in Figure~\ref{figure:experiment_acc}. Both forecasting models constructed by using the time-decayed reachability graph and the time-decayed transaction amount graph are able to achieve accuracy over 60\%, demonstrating the ability to correctly forecast transactions between accounts. However, the uncertainty of address life span and the evolving transaction pattern hinder further improvement in forecasting accuracy. The former would cause situations where one of the two accounts with frequent transactions disappears and the latter could result in cases where accounts in a small transaction community in the past may start transactions with new accounts recently. As shown in Figure~\ref{figure:experiment_f1}, we also achieve a reasonably high f-1 score. Again the results indicate that the proposed transaction forecasting model maintains good accuracy for predicting both the existence of transactions between accounts and the non-existent transactions between accounts.

\begin{table}[ht]
\centering
\scalebox{0.77}{
\small{
\begin{tabular}{|c|c|c|c|c|c|c|c|c|c|}
\hline
 & t1 & t2 & t3 & t4 & t5 & t6 & t7 & t8 & t9 \\ \hline
 static &	125 & 	506 &	1052 &	1539 &	1947 &	2280 &	2753 &	3262 &	3765\\ \hline
reachability & 10.5 & 22.9 & 34.7 & 45.8 & 58.1 & 69.5 & 81.3 & 92.6 & 105.2 \\ \hline
amount & 11.7 & 23.4 & 35.2 & 47.2 & 58.9 & 70.7 & 82.7 & 94.4 & 106.1  \\ \hline
\end{tabular}
}}
\caption{Runtime measurement of training node embedding for Bitcoin transaction graph. Time is measured in seconds.}
\label{table:runtime}
\vspace{-0.3cm}
\end{table}

The experiment shows that dynamic embedding of the Bitcoin transaction graph is always beneficial.
When only concerning if two nodes are connected or not without any time evolution information, the transaction forecasting performance of the baseline static graph is close to random guess, showing the strength of the constructed enhanced time-decay graphs. Since the static graph considers the embedding of all accounts at each time slot, Table~\ref{table:runtime} shows that the training time for embedding the two dynamic graphs is much shorter than embedding the static graph due to the ability of "forget" in dynamics graphs. The training time for different time-decaying graphs is approximately the same. The test time for all three graphs is approximately 0.7s.

Blockchain ledgers can grow very large over time. The Bitcoin blockchain currently requires around 200 GB of storage, and it doubles or triples the size when putting them into the memory for graph representation learning. Instead of mining over the entire history of Bitcoin transaction data, we choose the two subsets that represent two extreme cases that we want to study: sporadic and frequent, one at the beginning and the other at the latest time. We study the forecasting over the temporal partitions of the Bitcoin transaction data, between the two timeframes, over 9-10 years. First, we want to build the temporal sequences of transaction datasets, aiming to evaluate the effectiveness of our Bitcoin transaction forecasting system. By using the first dataset, we build a model to learn to predict the next in the sequence of our datasets. This will allow us to show how we utilize graph representation learning models to capture the temporal and spatial patterns of Bitcoin transactions over the span of the 9-10 years between the two timeframes. Second, we also want to utilize the temporal partitions of transaction data between the two timeframes over the 9-10 years to explore some general patterns. We report our findings in  Figure~\ref{figure:experiment_f1}, Figure~\ref{figure:experiment_acc}, and Table~\ref{table:runtime}. Our experimental evaluations were performed over two 100k-transaction datasets separated by 9-10 years. We use t1 to t9 in Figure 8 and Figure 9 as the set of  point-of-time in the temporal partitions of the Bitcoin-dataset. In fact, any dataset partition of transactions, occurred during the 9-10 years between the two periods represented by the two chosen subsets, are quite similar to either of the two, and thus the proposed system is directly applicable to them. Consider two accounts, say $A1$ and $A2$, have direct transaction relationship in the latest dataset, and account $A1$ also appeared in the earlier dataset, one can trace the temporal sequence of datasets over the 9-10 years to gain some understanding on how, when and through which other accounts that facilitate account $A1$ and account $A2$ to start transactions.  Since the impact of past transactions on each account decays differently, omitting historical transactions can lead to inaccuracy in prediction.

\section{Ensemble with Portfolio Selection}

In previous experiments, we observe that the behavior of the Bitcoin transactions is highly time-sensitive. Each of the time-decayed reachability graph and the time-decayed transaction amount graph has its own strength in transaction forecasting at different time periods during the data collection. Since the transaction pattern changes over time and the forecasting performance relies heavily on the data itself, no single method can outperform all others. Therefore, we provide a portfolio-based ensemble to decide which combo of forecasting models to use to reduce performance variance and maintain a stable yet competitive forecasting performance.

In an online decision setting, we iteratively choose transaction forecasting models constructed from embedding from different Bitcoin transaction features without knowledge of the future. For each pair of addresses $i=1,2,\cdots,$, the decision-maker chooses a forecasting model $f_k,1\le k \le K$ from the model set $\{f_1,...,f_K\}$ where $K$ denotes the size of the forecasting model set. Then, a cost function $L(f_k^t)$ is presented. When the outcome distribution is fixed, the performance measure of such online decision problem is defined as regret: the accumulated difference between the cost of the chosen decision and the best decision $L(f^{t*})$ in hindsight:
$$R^t=\sum \nolimits_t L(f_k^t)-\min \nolimits_f \sum \nolimits_t L(f^{t*}).$$
A good decision strategy for forecasting model selection would ensure that the regret converges fast to the optimal choice of the forecasting model as the number of game iterations grows. However, the underlying prediction outcome distribution may change in the highly dynamic Bitcoin transaction. For example, nodes with sparse connections tend to have more transactions in the past and may tend to stay inactive recently. Then, a good forecasting model for such nodes in the past may not be effective now and standard regret may not be the best measure of performance. Consequently, we extend the definition of regret to the maximum regret it achieves over any contiguous time interval:
$$\text{DR}^t=\sup \nolimits_{I=|r,s|\subseteq|T|}\{\sum \nolimits_{t=r}^s L(f_k^t)-\min \nolimits_f \sum \nolimits_{t=r}^s L(f^{t*})\}.$$

 \begin{algorithm}
\caption{Multiplicative Model Updates}
\begin{algorithmic}[1]
\STATE Let $S^t$ be the set of forecasting models at time $1\le t \le T$. Initialize $p^0=\frac{1}{K^0}$ where $K^0=|S^0|$. For any $t$, $p^t$ is a model selection distribution over forecasting model set $S^t$.
\FOR{t=1 in T do}
\STATE $\forall k \in S^t$, the choice of forecasting model is determined by the performance of the models in previous interval. The choice of the forecasting model in the current interval is computed as $f^t=\sum\nolimits_{k \in K^t}p^t f^t_k$.
\STATE when the forecasting cost $L(f^t)$ is observed, the model selection distribution is updated for $k \in S^t$:
$$p^{t+1}_k=\frac{p^t_k e^{-\alpha L(f^t_k)}}{\sum\nolimits_{j \in S^t}p^t_j e^{-\alpha L(f^t_j)}}$$
\STATE add new models. Set $\bar p_{k}^{t+1}$ to $\frac{1}{t+1}$ and for $i \neq t+1, \bar p_{i}^{t+1}=(1-(t+1)^{-1}) p_{i}^{t+1}$
\STATE remove low-quality models. Update $S^t$ to $S^{t+1}$ after adding new models. Then for all $i\in S^t$,
$$\hat p^{t+1}_k=\frac{p^{t+1}_k}{\sum \nolimits_{j \in S_{t+1}} p^{t+1}_j}.$$
\ENDFOR
\end{algorithmic}
\label{algo}
\end{algorithm}

\begin{figure*}[ht]
\begin{minipage}{0.49\linewidth}
 \centerline{\includegraphics[scale=.33]{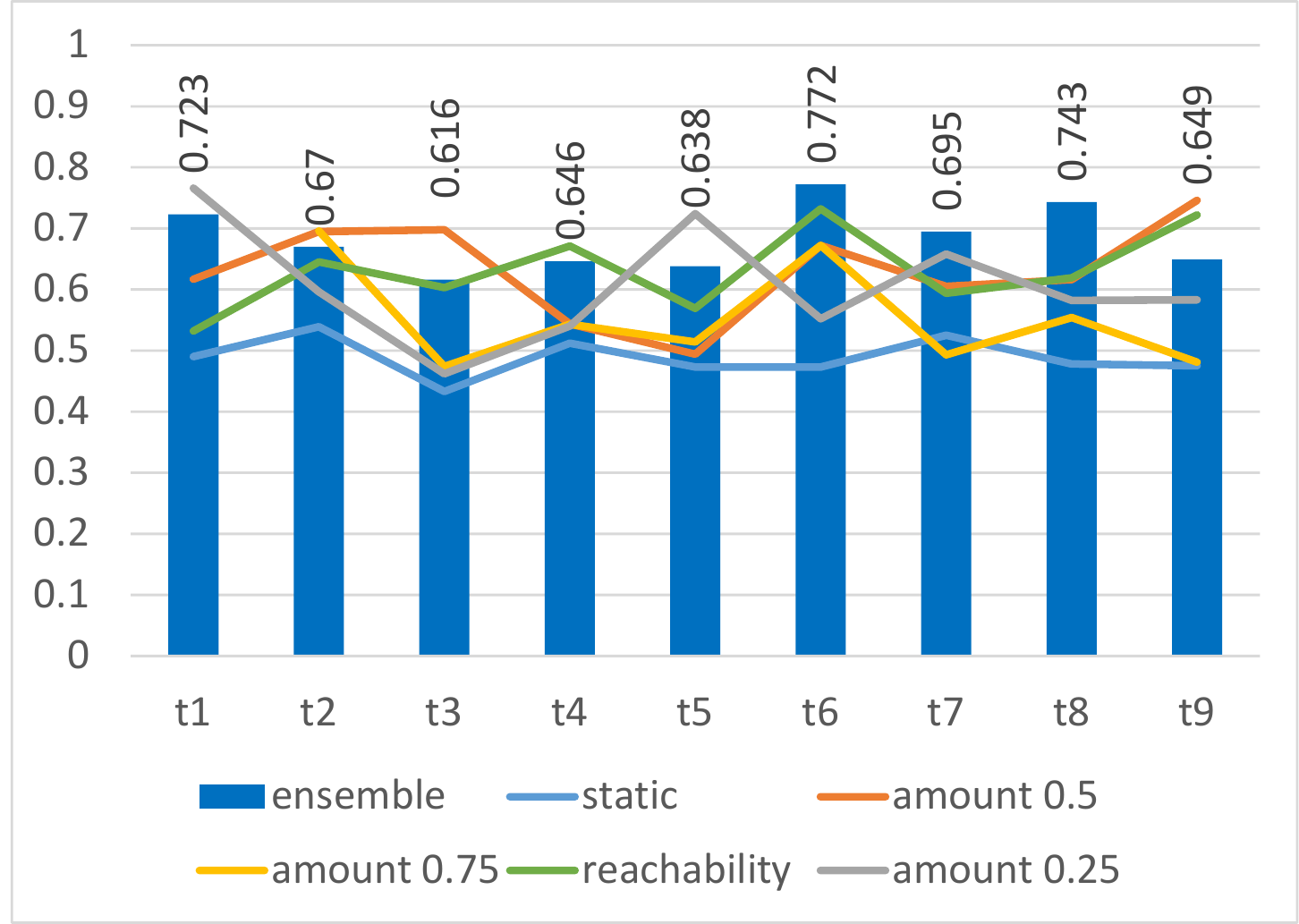}}
 \subcaption{Ensemble accuracy first 100k}
 %\vspace{-0.3cm}
 \label{figure:ens_acc_first}
\end{minipage}
\begin{minipage}{0.49\linewidth}
 \centerline{\includegraphics[scale=.33]{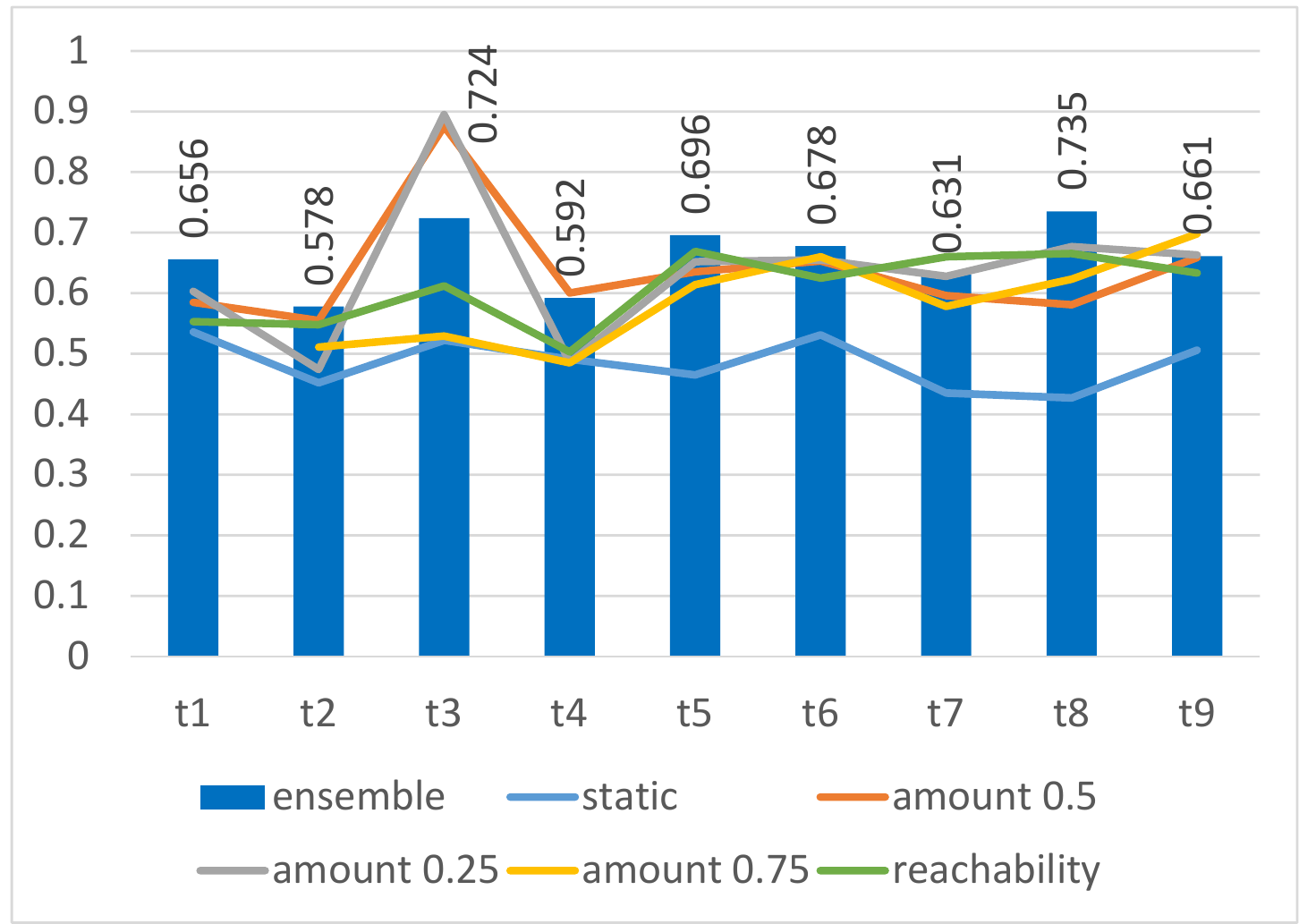}}
  \subcaption{Ensemble accuracy latest 100k}
   \label{figure:ens_acc_last}
\end{minipage}
\caption{Transaction forecasting accuracy using portfolio ensemble}
\label{figure:ensemble_acc}
\vspace{-0.3cm}
\end{figure*}

\begin{figure*}[ht]
\begin{minipage}{0.49\linewidth}
 \centerline{\includegraphics[scale=.33]{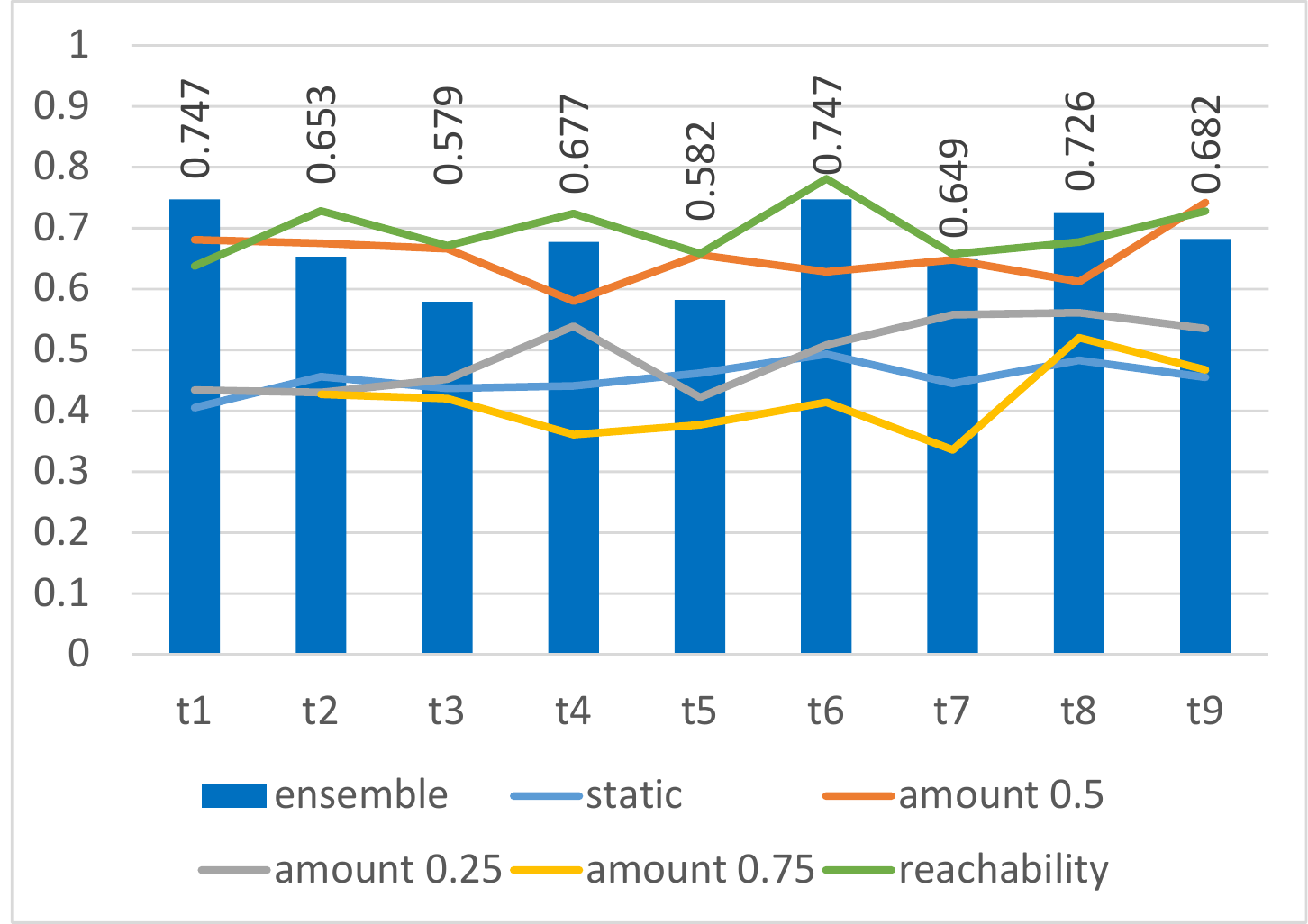}}
 \subcaption{Ensemble f1 first 100k}
 %\vspace{-0.3cm}
 \label{figure:ens_f1_first}
\end{minipage}
\begin{minipage}{0.49\linewidth}
 \centerline{\includegraphics[scale=.33]{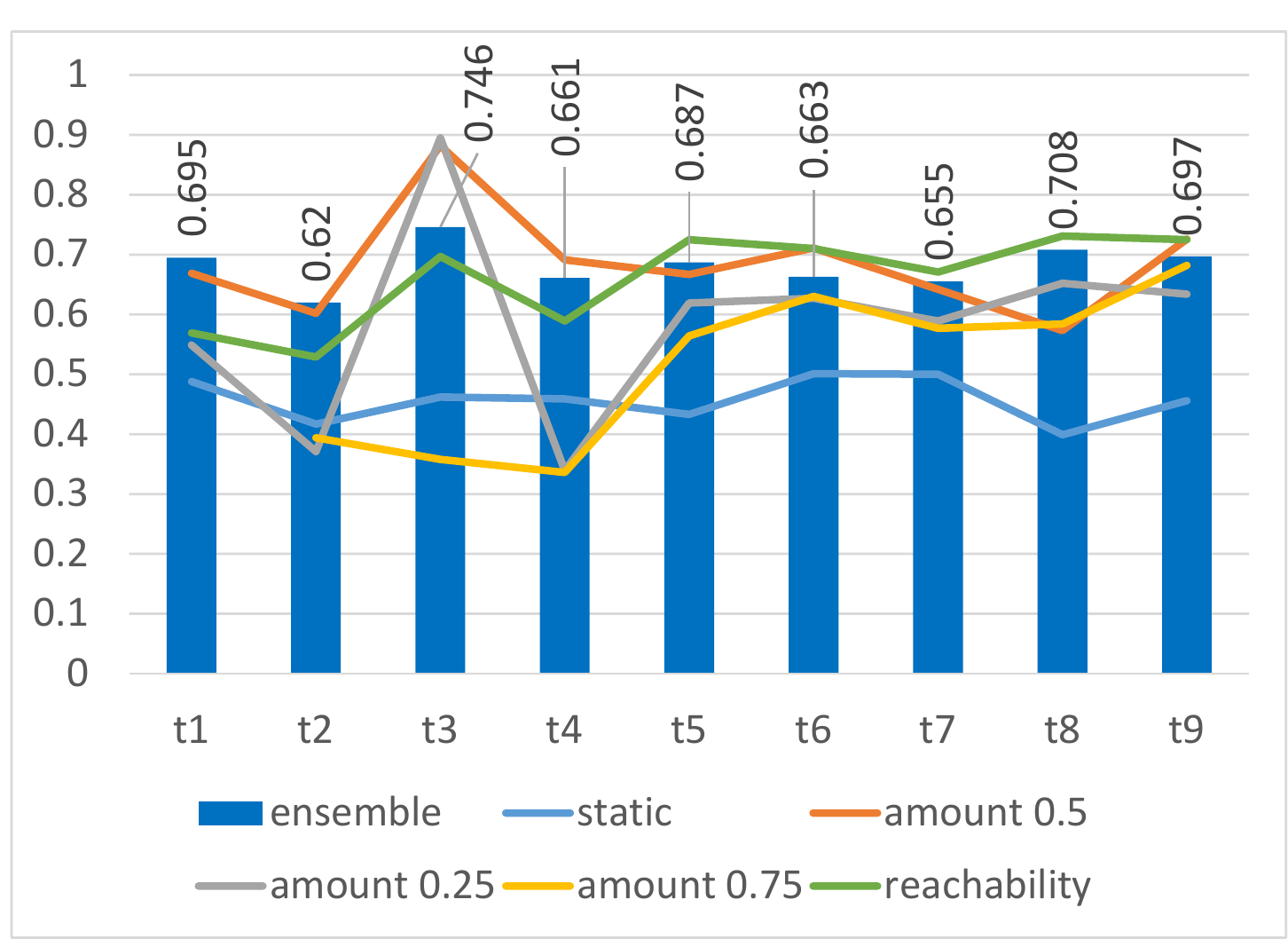}}
  \subcaption{Ensemble f1 latest 100k}
   \label{figure:ens_f1_last}
\end{minipage}
\caption{Transaction forecasting f1 using portfolio ensemble}
\label{figure:ensemble_f1}
\vspace{-0.3cm}
\end{figure*}

According to \cite{hazan2009efficient}, for exp-concave loss functions, an algorithm given $R^t$ regular regret in the fixed environment will have a $$\text{DR}^t \le R^t\log t + O(\frac{1}{\alpha}\log^2 t)$$ in the changing environment. We say a loss function is exp-concave if the function $e^{-L}$ is concave. Similar to the universal portfolio selection~\cite{cover2011universal}, we set the exp-concave loss function as $L(f)=-log(f*{r^t})$ where $r^t$ is a non-negative return vector which measures the ratio of the forecasting performance at $t$ to the forecasting performance at $t-1$ for the corresponding model. To choose amongst different forecasting models, we apply a well studied Multiplicative Weights method~\cite{arora2012multiplicative} and name the model selection procedure at each forecasting period interval as Multiplicative Model Updates. Algorithm~\ref{algo} gives a sketch of the model selection idea. Line~3 choose the forecasting model according to the forecast performance of the models in the ensemble in previous interval. The choice of the forecasting model in the current interval is computed as $f^t=\sum\nolimits_{k \in K^t}p^t f^t_k$. Line~4 indicates that when the forecasting cost $L(f^t)$ is revealed, we will use the observed forecasting results to update the model selection distribution of model $k$: $p^{t+1}_k=\frac{p^t_k e^{-\alpha L(f^t_k)}}{\sum\nolimits_{j \in S^t}p^t_j e^{-\alpha L(f^t_j)}}$. Line~5 adds new models into the ensemble with $\bar p_{k}^{t+1}=\frac{1}{t+1}$ and set $\bar p_{i}^{t+1}=(1-(t+1)^{-1}) p_{i}^{t+1}$ for $i \neq t+1$. Line~6 remove the models that show poor forecasting performance in previous intervals and update $S^t$ to $S^{t+1}$ after adding new models. Then for all $i\in S^t$,
$\hat p^{t+1}_k=\frac{p^{t+1}_k}{\sum \nolimits_{j \in S_{t+1}} p^{t+1}_j}$. The low quality in Line 6 refers to the model with low forecast accuracy. At each forecasting time, the choice of the forecasting model is determined by the performance, i.e., the prediction accuracy of the models in the previous interval.  We remove some models of low quality and add new models to capture and to deal with the dynamics of the Bitcoin graph. Our initial results are conducted by removing the model with the lowest forecast accuracy and add one by setting its model selection distribution according to Algorithm 1 line 5.

Following the Multiplicative Model Updates algorithm, we construct a portfolio selection ensemble to decide which forecasting model to use at each query.
We generate the working model set by constructing forecasting models using different embedding features. The embedding is generated from graphs using different time-decay factors and has different starting points. The idea of using different decay factors is originated from the observation that the life span of addresses is very different.
While each feature graph is at best in capturing some kinds of transaction patterns, e.g., connectivity, transaction amount, or some hidden features, which feature graph best preserves transaction patterns is highly dependent on data. A continuously active address would require a long decay factor while a one-time transaction address should have a short decay factor. Since the best transaction pattern-preserving scale can not be known beforehand, we inject multiple time-decay factors to produce node embedding and construct transaction forecasting models. To be specific, we apply time-decay factors of 0.25 and 0.75 in addition to 0.5 on the transaction amount graph. The idea of choosing different starting points is based on the existence of local transaction patterns. As illustrated in Figure~\ref{figure:bitcoin_amount}, sender-receiver pairs with a large number of Bitcoins are more frequent in some periods than in other periods.

The accuracy and f1-score measurement in Figure~\ref{figure:ensemble_acc} and Figure~\ref{figure:ensemble_f1} confirms our analysis that different time-decay factors in the transaction amount graph would capture different transaction patterns and no extracted transaction feature would always outperform other features in the transaction forecasting task. Due to the high Bitcoin transaction dynamics, we observe that the graph with a long time-decay factor is less efficient as both the accuracy and f-1 score of the forecasting model constructed with a long time-decay factor graph are relatively lower when compared with forecasting models built upon other graphs. Although the ensemble cannot always maintain a forecasting performance as best as the best single model in the working model set, the results
indicate that the ensemble would ensure us not to choose the worst decision-maker sequentially. Meanwhile, when each single forecasting vector may suffer from low (or high) accuracy and high (or low) f-1 score, the portfolio-based ensemble would keep both accuracy and f-1 score competitive.

\section{Conclusion}

We have presented DLForecast $-$ a Bitcoin transaction forecast system, which leverages deep neural networks to learn Bitcoin transaction network representations. This paper makes three unique contributions. First, we analyzed the Bitcoin transaction data by exploring their transaction-based connectivity patterns and their transaction amount patterns. Second, we constructed a time-decayed reachability graph and a time-decayed transaction pattern graph to extract spatial and temporal features of Bitcoin transaction dynamics. Third but not the least, we learn Bitcoin transaction patterns through node embedding by mapping each of the constructed graphs into a low-dimension representation vector space. Through iterative network embedding training, we build a deep neural network-based Bitcoin transaction forecasting model, which is capable of making predictions on the transaction patterns between user accounts based on historical transactions and the built-in time-decaying factor. Evaluated on real-world Bitcoin transactions, we showed that our spatial-temporal forecasting model is efficient with fast runtime, effective with forecasting accuracy over 60\%, and it improves the prediction performance by 50\% when compared to the forecasting model built on the static graph.

In addition to deploying DLForecast for Bitcoin transaction forecasting, the proposed  system can also be used to detect and identify certain interesting transaction behaviors, e.g., some accounts may exhibit short term absence or presence in their history of transactions. DLForecast can additionally be used to monitor legitimate transactions and identify illicit actors in the crypto space. Even though Bitcoin transactions include no personally identifiable information about users, such as names, addresses, or social security numbers, the dynamic graphs constructed by the DLForecast system can be used to connect multiple transactions to the same account. Thus, such dynamic graphs can be utilized for identifying certain behavior patterns of a single address, such as long term transaction of a small amount of Bitcoins and a sudden large amount transaction, and for associating such transaction behavior with some real-world events or timeline, which may assist the law enforcement to track those transactions made by illicit actors (dark marketplaces, ransomware operators, fraudsters) and to identify those transactions made by legitimate actors (e.g., regulated exchanges, merchants, wallet services). Another interesting utility of DLForecast is to look into those cases where a transaction happens when the forecasting model predicts such a transaction as unlikely to happen for a given period. Although DLForecast is developed for analyzing and predicting Bitcoin transactions, the proposed system and algorithms developed can be applied to a range of cryptocurrencies and blockchain-based assets, such as those for storing financial records or any other data where an audit trail is required because every change is tracked and permanently recorded on a distributed and public ledger. The proposed system can help reducing compliance costs and monitoring and detecting criminal or illegal activities.
%% However, Blockchain ledgers can grow very large over time. The Bitcoin blockchain currently requires around 200 GB of storage, making it hard for deploying current machine learning techniques for data mining. For future work, we would like to scale the analysis to the entire dataset by designing a more scalable node embedding techniques that can feed the large Bitcoin graph into memory or can distributedly process the graph.

\section*{Acknowledgment}
The first author thanks the opportunity of the 12-week working experience at IBM T. J. Watson Research Center in Summer 2019 with the group led by Donna N Dillenberger. This work is partially sponsored by NSF CISE grant 1564097 and an IBM faculty award.

%\clearpage

\bibliographystyle{IEEEtran}
\bibliography{bare_jrnl_compsoc}

\begin{IEEEbiography}
[{\includegraphics[width=1in,height=1.25in,clip,keepaspectratio]{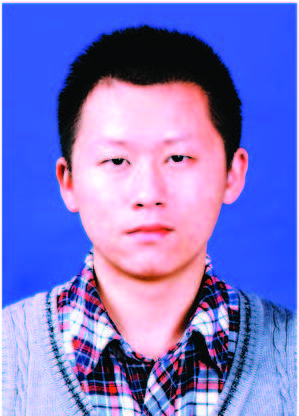}}]{Wenqi Wei}
is currently pursuing his Ph.D. in the School of Computer Science, Georgia Institute of Technology, where he is advised by Prof.~Ling Liu. He received his B.E. degree from the School of Electronic Information
and Communications, Huazhong University of Science
and Technology. His research interests include data privacy, security, machine learning, and big data analytics.
\end{IEEEbiography}

\vspace{-20pt}

\begin{IEEEbiography}
[{\includegraphics[width=1in,height=1.25in,clip,keepaspectratio]{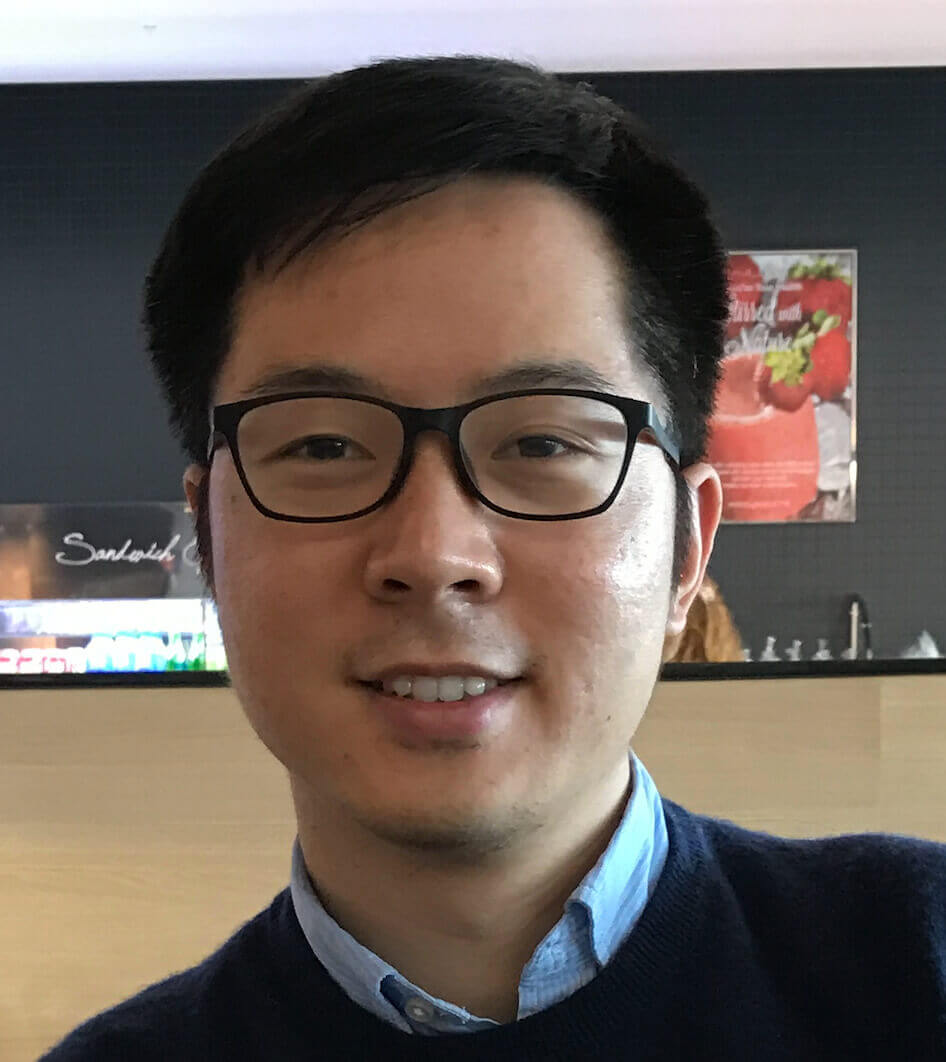}}]{Qi Zhang}
Qi Zhang is a Research Staff Member at IBM Thomas J. Watson Research Center, received his Ph.D. degree in computer science from Georgia Institute of Technology (USA) in 2017. His research interests include blockchain systems, cloud computing, big data, and deep learning systems, and published in major refereed journals, such as IEEE TC, IEEE TSC, ACM CSUR, IEEE Blockchain NewsLetter, and conferences, such as VLDB, Blockchain, IEEE ICDCS, SuperComputing, ICWS, CLOUD, and
VLDB, HPDC.
\end{IEEEbiography}

\vspace{-20pt}

\begin{IEEEbiography}
[{\includegraphics[width=1in,height=1.25in,clip,keepaspectratio]{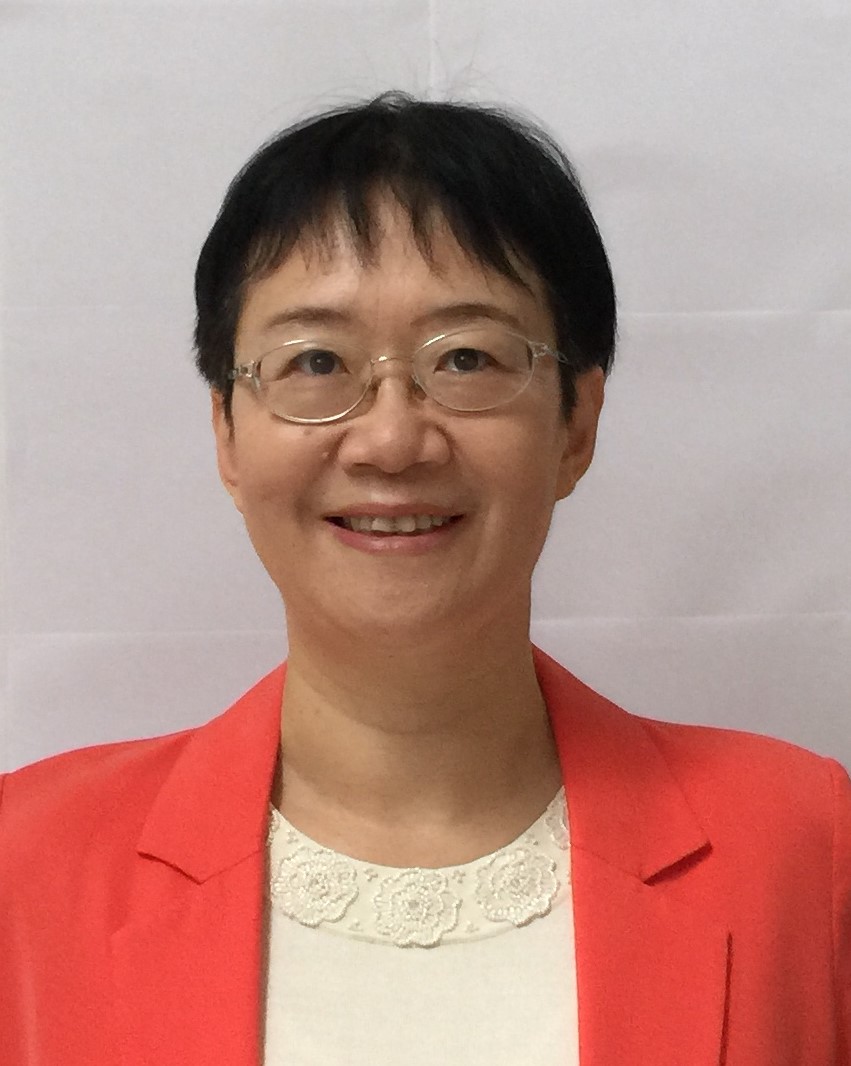}}]{Ling Liu}
is a professor in the School of Computer Science, Georgia Institute of Technology. She directs the research programs in the Distributed Data
Intensive Systems Lab (DiSL). She is an elected IEEE fellow, a recipient of the IEEE Computer Society Technical Achievement Award in 2012, and a recipient of the best paper award from a dozen of top venues, including ICDCS 2003, WWW 2004, 2005 Pat Goldberg Memorial Best Paper Award, IEEE Cloud 2012, IEEE ICWS 2013, ACM/IEEE CCGrid 2015, and IEEE Symposium on BigData 2016. In addition to serving as the general chair and PC chair of numerous IEEE and ACM conferences in data engineering, she has served on the editorial board of over a dozen international journals. Her current research is primarily sponsored by NSF, IBM, and Intel.
\end{IEEEbiography}

\end{document}